\documentclass[acp,manuscript]{copernicusarxiv}

\begin{document}

\title{The ocean fine spray}


\Author[1,2]{Alfonso M.}{Ga{\~n\'a}n-Calvo}

\affil[1]{Departamento de Ingenier{\'\i a} Aeroespacial y Mec{\'a}nica de Fluidos, ETSI, Universidad de Sevilla, Camino de los Descubrimientos, 41092 Sevilla, Spain}
\affil[2]{ENGREEN, Laboratory of Engineering for Energy and Environmental Sustainability, Universidad de Sevilla, Camino de los Descubrimientos, 41092 Sevilla, Spain}




\correspondence{amgc@us.es}

\runningtitle{The ocean fine spray}

\runningauthor{Ga{\~n\'a}n-Calvo}

\received{}
\pubdiscuss{} 
\revised{}
\accepted{}
\published{}


\firstpage{1}

\maketitle

\begin{abstract}
A major fraction of the atmospheric aerosols come from the ocean spray originated by the bursting of surface bubbles. A theoretical framework that incorporates the latest knowledge on film and jet droplets from bubble bursting is here proposed, suggesting that the ejected droplet size in the fine and ultrafine (nanometric) spectrum constitute the ultimate origin of primary and secondary sea aerosols through a diversity of physicochemical routes. In contrast to the latest proposals on the mechanistic origin of that droplet size range, when bubbles of about 10 to 100 microns burst, they produce an extreme energy focusing and the ejection of a fast liquid spout whose size reaches the free molecular regime of the air. Simulations show that this spout yields a jet of sub-micrometer and nanometric scale droplets whose number and speed can be far beyond any previous estimation, overcoming by orders of magnitude other mechanisms recently proposed. The model proposed  can be ultimately reduced to a single controlling parameter to predict the global probability density distribution (pdf) of the ocean spray. The model fits remarkably well most published experimental measurements along five orders of magnitude of spray size, from about 5 nm to about 0.5 mm. According to this proposal, the majority of ocean aerosols would have their extremely elusive birth in the collapsing uterus-like shape of small bursting bubbles on the ocean surface.
\end{abstract}

\begin{figure*}[!t]%
\centering
\includegraphics[width=120mm]{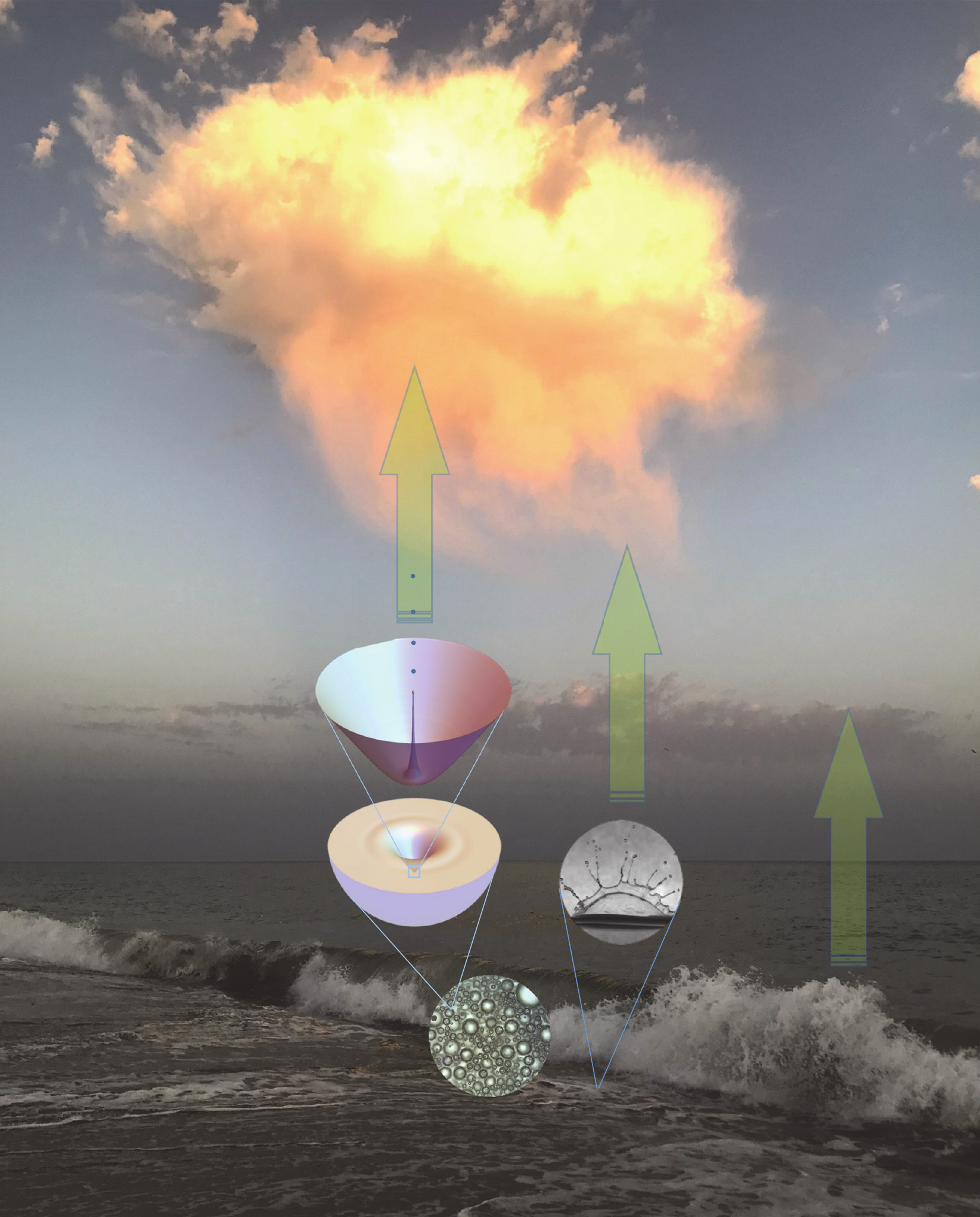}
\caption{Marine environment from which a continuous flow of primary aerosols is generated, schematically indicating the basic mechanisms of spray formation: turbulent fragmentation (spume drops) and bubble bursting (film and jet droplets). Jet droplets from microscopic bursting could be not only much more numerous than initially thought, but also their extremely small size, in the range of free molecular flow, and their astonishing ejection speed could make them the main source of aerosols in the atmosphere. (Original photograph by the author made at La Antilla, Huelva, Spain, during sunset, pointing Southwest, August 2020. The sun is illuminating the cloud from the right while the beach is in twilight)}
\label{f1}
\end{figure*}

\introduction  
The basic mechanism of fine seawater fragmentation, essential for primary ocean aerosol production, is the bursting of bubbles produced by breaking waves (figure \ref{f1}). From ultrafine to coarse size, the bubble bursting spray leads to the nascent sea spray aerosols (nascent SSA, or nSSA), and primary and secondary marine aerosols (PMA and SMA) whose composition and transport depends on the size of the initial droplets and their bio-physicochemical route along their lifetime \citep{Bates2012,Schmitt-Kopplin2012,Bertram2018,Brooks2018,Trueblood2019,Mayer2020,Mitts2021,Deike2022,Angle2021,Cornwell2021}.
These aerosols determine vital self-regulating planetary mechanisms from the water cycle dynamics to the atmospheric optical thickness and planetary albedo via aerosol micro-physics (cloud nucleation, chemical reactions, and catalyzed condensations) with a dominant impact on the radiant properties of atmosphere and global climate, among other primary effects.
Current literature \citep{Cochran2017,Wang2017,Brooks2018,Mayer2020,Cornwell2021,Angle2021,Liu2022} describes in great detail the chemical composition of ocean aerosols and their dependence on ambient (temperature, wind speed, biological activity, latitude etc.), geographical or lifetime parameters. However, knowing with precision their ultimate origin is often a hopeless task: their generation usually entail extremely elusive phenomena. Indeed, one of the smallest scale, most elusive, extraordinarily fast, yet ubiquitous phenomena of continuous media is the emission of droplets from the bursting of small bubbles at the surface of water.

Incomplete or incorrect causal attributions in science are strongly correlated with the limitations of instruments and tools able to observe the very large or very small spatial and temporal scales \citep{Penrose2000}, limitations which often lead to periods of stymied progress. However, the assembly of indirect evidences from multiple sources and methods \citep{Wang2017,Jiang2022} is the usual process of advancement. This work presents a thorough revision of the physics of bursting bubbles and their associated statistics, and proposes a global statistical model to describe the size distribution of the average ocean spray.

Two basic mechanisms are responsible of this droplet emission: bubble film breakup \citep{Lhuissier2012,Jiang2022} and jet emission \citep{Worthington1908,Woodcock1953,Cipriano1981}. The super-micron size range is comprised by wet aerosols and its presence is fundamentally reduced to marine and coastal regions \citep{Boyce1954}. In contrast, the sub-micron aerosol size range encompassing the Aiken (10 to 100 nm) and the accumulation (100 nm to about 1 micron) modes \citep{Pohlker2021}, where cloud condensation nuclei (CCN) and ice nucleation particles (INP) are included, is present everywhere in the atmosphere up to the stratospheric layers.

The origin of the sub-micron aerosol population has been historically attributed to the smallest size range of film breakup droplets \citep{Cipriano1981,Resch1986,Wu2001,Prather2013}, an idea that has not been challenged until the recent work of \cite{Wang2017}. These authors were probably the first ones quantitatively demonstrating that jet droplets could be more important than previously thought. They imputed the differences found in the chemical composition of their collected aerosols to the potentially different origin of the liquid coming from either the bubble film or the emitted jet. However, that difference could also be imputed to the changing relative size of the smaller bursting bubbles compared to the surface microlayer thickness \citep{Cunliffe2013}.

In a recent work, \cite{Berny2021} have also pointed to jet droplets as the potential cause of aerosols in the range down to 0.1 $\mu$m with maximum number probability density, according to these authors, around 0.5 $\mu$m. Indeed, dimensional analysis and up-to-date models reveal that jet drops from bursting seawater bubbles with sizes from about 15 to 40 microns can yield at least tens of submicron jet droplets with sizes down to about 200 nm \citep{Brasz2018,Berny2020,GC2021}. Moreover, \cite{Berny2022} observed the ejection of secondary jet droplets, sensitive to initial conditions of the bursting process, much smaller than the primary ones.

A bold proposal has been very recently published \citep{Jiang2022} to explain the submicron SSA origin from film droplets: the film flapping mechanism \citep{Lhuissier2009}. This mechanism aims to complement the film bursting described by \cite{Lhuissier2012} for the complete description of the spray size distribution, disregarding jet droplets. The authors provide probably the most comprehensive collection of experimental data on collective bubble bursting so far together with \cite{Neel2021}, to the best of our knowledge, including a highly valuable statistical disaggregation by both bubble and droplet size, while the recent works of Berny et al. \citep{Berny2021,Berny2022} are probably the best sources of numerical information on collective bubble jetting.


Bubble bursting (BB) is a common phenomenon of liquid phase. However, liquids with a low viscosity and relatively large surface tension exhibit special features. Consider the average density, viscosity and surface tension of seawater at the average surface temperature of ocean (15$^o$C): $\rho=1026$ kg m$^{-3}$, $\mu=0.00122$ Pa$\cdot$s, and $\sigma=0.0743$ N m$^{-1}$ respectively. The best reference measures to describe the physics of BB are the natural scales of distance, velocity and time defined as $l_\mu= \mu^2/(\rho \sigma)=19.5$ nm, $v_\mu=\mu/\sigma=61$ m/s, and $t_\mu=\mu^3/(\rho \sigma^2)=0.32$ ns for seawater. These scales allow the rationalization and comparison of the different extremely rapid mechanisms of droplet generation. Using all data provided by \cite{Berny2021,Neel2021,Berny2022,Jiang2022} among other valuable information and data resources, jet and film droplet generation from seawater are here exhaustively revised under current available theoretical proposals \citep{GC2021,Jiang2022} and experimental evidences. Disaggregated data and detailed experimental description in \citep{Jiang2022} allow reliable statistical resolution of ambiguities in the origin of droplets in the micron and submicron range.

In this work, a comprehensive physical rationale for the spray generation from the ocean is proposed incorporating all mechanisms (film bursting, film flapping, and jetting) into a global model. The associated statistics and physical models are reduced to closed mathematical expressions fitted to the existing supporting data. The expressions obtained are subsequently integrated into the general statistical model of the ocean spray proposed. This model predicts the number concentration (probability density function) of the average oceanic spray size. The proposed model is compared with an extensive collection of ocean aerosol measurements. To do so, the measured particle size (diameter) is converted to the presumed originating droplet radius, assuming evaporation in the majority of cases, and some degree of condensation for the smallest aerosol size ranges. Given the wide range of sizes considered, the impact of the accuracy of this conversion is expected to be marginal. Indeed, the surprising agreement to experimental measurements found along five orders of magnitude of droplet radii $r_d$ would provide a strong confidence on the proposed description. The quantitative agreement suggests that the origin of both primary SSA and SMA would definitely be bubble jetting, with a minor contribution of film flapping droplets \citep{Jiang2022}.

\section{Droplet statistics per bursting event}

\subsection{Film droplets: statistics}

\begin{figure}[!t]%
\centering
\includegraphics[width=0.75\textwidth]{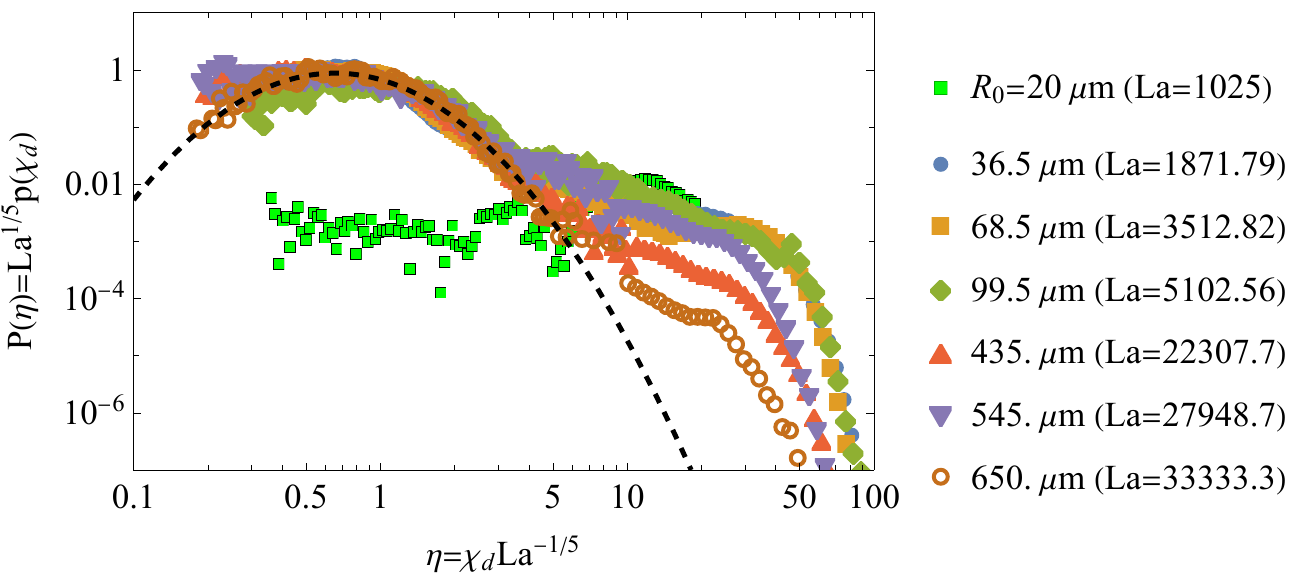}
\caption{Probability density function $P(\eta)=$La$^{1/5}p(\chi_d)$ for the droplet radii $\chi_d=r_d/l_\mu$ scaled as a non-dimensional variable $\eta=\chi_d$La$^{-1/5}$. For appropriate fitting purposes to a lognormal(thick dashed line), the experimental pdfs \citep{Jiang2022} are multiplied by 2 to approximately compensate the number contribution for sizes $\eta\gtrsim 4$.}
\label{f2}
\end{figure}

The generation of droplets in the micron-size and above from the disintegration of the bubble cap film rims is exhaustively described by \cite{Lhuissier2012} and references therein. On the other hand, the physics of the film flapping proposal for the production of submicron droplets by \cite{Jiang2022} is described in detail in that work. Their experimental results allow a detailed disaggregated statistical analysis of droplet generation per bubble size. From these data, we have found a very useful scaling law for our purposes that collapses their experimental probability distributions on a lognormal distribution for $r_d\lesssim 0.5$. The universal distribution found has a mean value $\langle \chi_d\rangle = \langle r_d\rangle /l_\mu=1.1$La$^{1/5}$ and variance $\nu = 0.5$, for equivalent bubble radii from $R_o\sim 35 \mu$m to 0.7 mm as a function of the Laplace number, as shown in figure \ref{f2}.

Note that \cite{Jiang2022} use what they call {\sl the radius of curvature of the cap} as the equivalent bubble radius $R_o$, which according to their calculations is approximately twice the radius of a sphere with the same volume of the bubble for small Bond numbers Bo=$\rho g R_o^2/\sigma$. For bubbles larger than about 0.5 mm, they use Toba's correction \citep{Toba1959}. Given that the general understanding on the bubble radius $R_o$ in the literature \citep{Spiel1995,Duchemin2002,Brasz2018,Berny2020,GC2021} is that of the equivalent volume sphere, we use this latter value here.

\begin{figure}[!t]%
\centering
\includegraphics[width=0.75\textwidth]{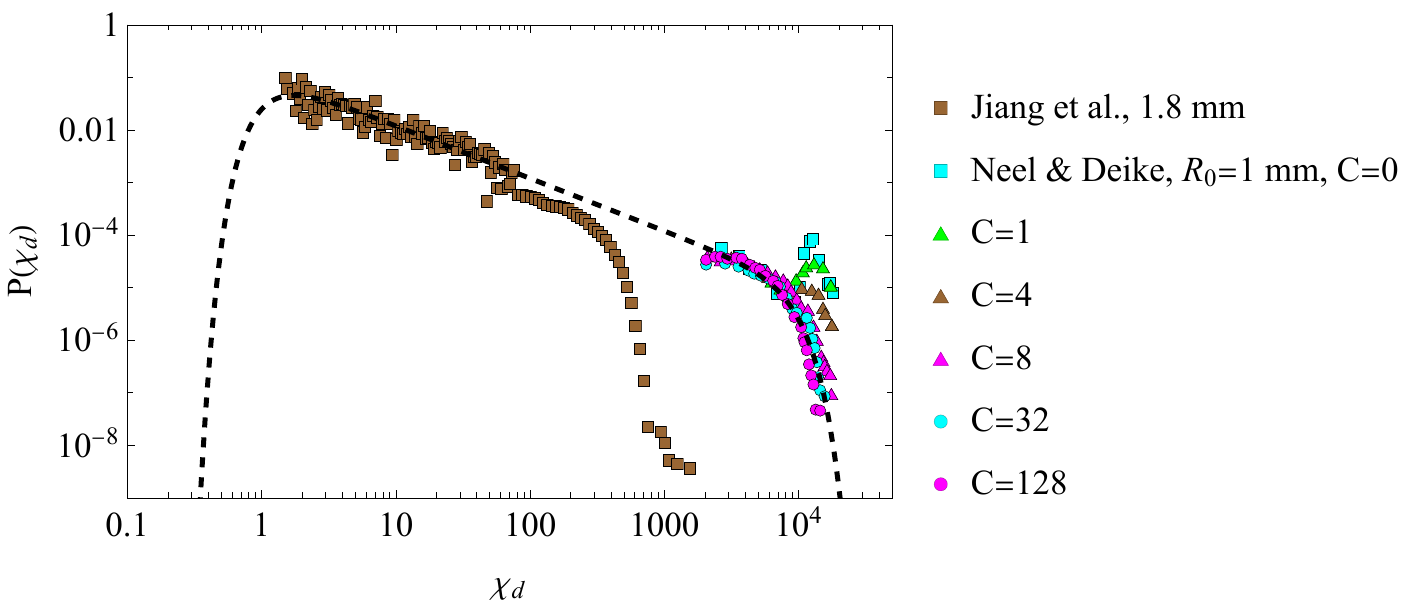}
\caption{Probability density function $P(\chi_d)$ for the droplet radii $\chi_d=r_d/l_\mu$ fitted to the experimental data of \citep{Jiang2022} for $R_o>$ 1 mm and of \citep{Neel2021} for different surfactant concentrations. The number densities from \citep{Neel2021} are appropriately scaled to collapse along the ordinates.}
\label{f3}
\end{figure}

As \cite{Neel2021} have recently pointed out, there is a very significant difference between the bulk and the surface bubble size distribution that eventually bursts for (i) clean seawater and bubbles around 1 mm and larger, and (ii) probably when the residence time at the surface before bursting is long enough to allow accumulation and coalescence \citep{Shaw2021}. This seems to be the case of the average bubble size $R_o=20$ reported by Jiang et al. in their supplementary information \citep{Jiang2022} which, incidentally, approximately coincides with the critical Laplace number La$_c$ described in \cite{GC2021}. This could explain the wide distribution of droplet sizes measured in that case, in line with the results of \cite{Neel2021}. These latter results are collected and displayed together with the data of \cite{Jiang2022} for $R_o$ larger than 1 mm in figure \ref{f3}.

The experimental setup and measurement equipment used in \cite{Jiang2022} might not allow to precisely determine the number concentration of droplet radii about ten microns and above due to impacting and settling, which would explain the fast decay for $\chi_d \gtrsim 350$. In contrast, the direct optical measurements of \cite{Neel2021} reliably cover droplet sizes up to 0.4mm. These latter authors find two types of droplets (see figure \ref{f3}) that can be attributed to film breakup (the collapsing data independent of the surfactant concentration $C$) or jetting (the peaks around 0.2 mm), which would agree with jet droplet size predictions \citep{GC2017,GC2021,Berny2022}. With these considerations in mind, an ensemble pdf can be constructed and fitted to the experimental data after the appropriate scaling of the probabilities reported by \cite{Neel2021}. The data are remarkably well fitted to a generalized inverse Gaussian distribution as:
\begin{equation}
P(\chi_d)=\frac{\beta (\langle \chi_d\rangle /\chi_d)^{-1}}{2 K_0(\gamma)}\exp\left(-\gamma\left( (\langle \chi_d\rangle /\chi_d)^\beta+(\chi_d/\langle \chi_d\rangle)^\beta\right)/2\right),
\label{GIG}
\end{equation}
with $\langle \chi_d\rangle = 10^2$, $\gamma=5\times 10^{-5}$, and $\beta=2.4$.

From this study and entirely attributing the fitted pdfs to micron- and submicron film droplets, one would conclude that (i) the droplet generation by film flapping will be distributed according to a lognormal for the nondimensional variable $\eta=\chi_d$La$^{-1/5}$, which reflects a reasonable dependency on the bubble size, while (ii) the more `` classical'' film rim fragmentation \citep{Lhuissier2012} would yield droplets distributed according to a generalized inverse Gaussian as (\ref{GIG}) independently of La.

\subsection{Jet droplets: statistics}\label{sec3}

\begin{figure}[!t]%
\centering
\includegraphics[width=0.75\textwidth]{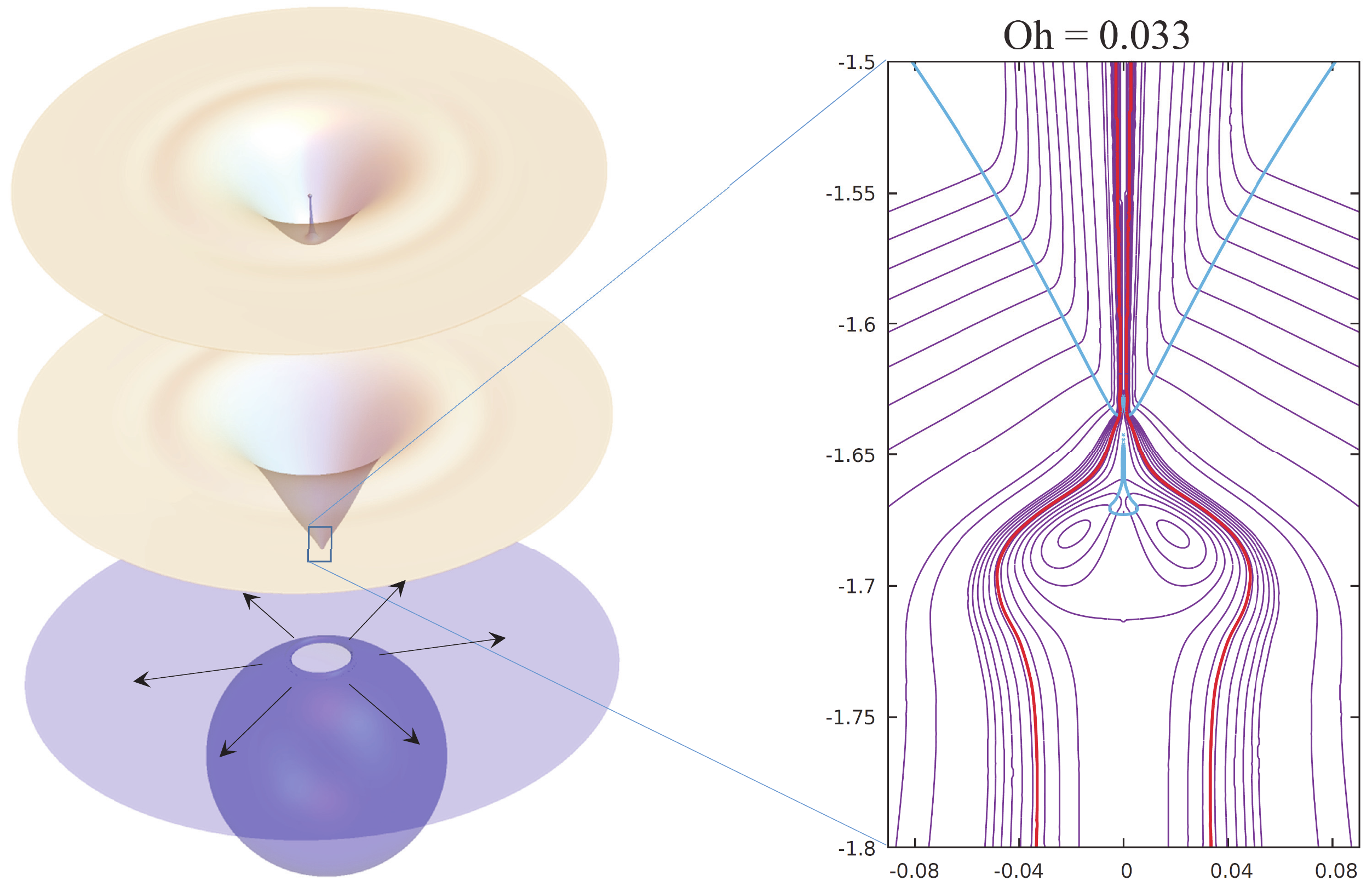}
\caption{(Left) Mechanism of bursting from an initially nearly spherical bubble at the surface of a liquid, and (Right) Streamlines (blue thin lines) of the liquid and air flow around the point of collapse for an instant about $t-t(0)=10^{-5} t_o$, where $t_o=\left(\rho R_o^3/\sigma\right)^{1/2}$, from Ga{\~n\'a}n-Calvo and L{\'o}pez-Herrera \citep{GC2021}. Oh is the Ohnesorge number, related to the Laplace number La = $R_0/l\mu$ as Oh = La$^{-1/2}$. The cyan thick line is the free surface depicted in the right panel. The streamline passing by the point where the vertical velocity of the surface is maximum is highlighted as a thick red curve (see text).}
\label{f4}
\end{figure}

Figure \ref{f4}(left) shows three instants of the bursting of a small, nearly spherical bubble at the surface of a liquid: (i) right after the puncture of the thin liquid film, (ii) when the bottom collapses into a nearly conical shape, producing the beginning of ejection, and (iii) when the first droplet is about to be ejected. Consider also the pattern of streamlines at the beginning of ejection (figure \ref{f4} right, from \cite{GC2021}) at the bottom of the cavity: this pattern indicates the origin of the liquid ejected as droplets.

Physical similarity is especially useful for investigating jetting from BB, since it is in principle (assuming negligible dynamical effects of the environment) a conceptually simple biparametric dimensional problem that can be studied in great detail from both experimental and theoretical approaches \citep{Duchemin2002,Walls2015,Ghabache2016,GC2017}. The two main parameters are the Laplace number La = $\rho \sigma R_o/\mu^2$ (or the alternative Ohnesorge number Oh = La$^{-1/2}$) and the Bond number Bo = $\rho g R_o^2/\sigma$. Experiments show that there is a critical Laplace number La$_c$ for which the ejected liquid spout reaches a minimum size with maximum ejection speed \citep{Duchemin2002,Walls2015,Ghabache2016,GC2021}. For La numbers around La$_c\simeq 1100$ (see figure 4, main text, for La = 918.3 or Oh = 0.033), the influence of density and viscosity ratio with the outer gas environment may become not only noticeable but also crucial to determine these minimum ejected droplet size and maximum speed \citep{GC2021}.

When seawater (and liquid water in general) is involved, though, the phenomenon is not directly observable in the La ranges around La$_c$ due to the smallness of $l_\mu$ and a direct experimental assessment of physical models is not possible. In effect, when gas bubbles from tens to hundreds of micrometers burst at a {\sl water} free-surface, a large numerical fraction of the emitted droplets lies out of the observable range, despite previous efforts by Lee et al. \citep{Lee2011}, who showed the elusive latest stages of jetting from a 45 $\mu$m bursting bubble using X-ray phase-contrast imaging. Even the most precise measuring instruments have limitations concerning the size, speed or temporal measurability of samples from these ejections. This is because the natural scales of seawater, distance $l_\mu=\mu^2/(\rho \sigma)$ (about 20 nm) and time $t_\mu=\mu^3/(\rho^2\sigma)$ (about 0.32 ns), are involved in the extreme ejection phenomena for bubble radii $R_o$ around La$_c l_\mu$, which are far beyond current optical and imaging instruments. The complexity of the problem is aggravated because those scales are, as subsequently shown, comparable or far below the scales of free molecular flow of the surrounding atmosphere at standard conditions, and hence the parametric dependency on the density and viscosity ratios with the environment become meaningless.

The different initial conditions of the bursting process and the numerical precision used in simulations may produce a significant variability around a critical La number, La$_c \simeq 1100$  (corresponding to a critical Ohnesorge number Oh$_c\simeq 0.03$ \cite{GC2021}). Interestingly, the data series from each source can be independently and accurately fitted by different $\alpha_1$ values, keeping the same $k_R$ and $\alpha_2$. This fitting parameter $\alpha_1$, which measures the relative magnitude of the surface tension pressure to produce the initial droplet, compared to the dynamic pressure, plays a determining role when La $\simeq$ La$_c$ (see expression (2), main text). Note that the data in figure 5, main text, correspond to the radius of the {\sl first} ejected droplet, $R$.

However, the emission lasts during times comparable to, or longer than the capillary time $t_o=\left(\rho R_o^3/\sigma\right)^{1/2}$, much larger than the time of formation of the first droplet at the front of the issuing liquid ligament, $t_c=\left(\rho R^3/\sigma\right)^{1/2}$ since $R\ll R_o$. Thus, the high speed liquid ligament has a long time to elongate and disintegrate into a large number of droplets with a variety of radii $r_d$ \citep{Berny2022}. In these conditions, the density and viscosity of the outer medium can dramatically alter the breakup of the spout into droplets.

This was demonstrated reducing the viscosity and density of the outer environment one order of magnitude (see figure 6, main text, in \cite{GC2021}), which keeps the high velocity of the jet front for a longer time. Indeed, the bubble radius corresponding to La$_c$ (minimum $r_d$) is $R_o\simeq 20\,\mu$m. This would lead to drop radii well below the mean free path of gas molecules of the environment, and ejection speeds above their average molecular speed. Hence, the values of the fitting constant $\alpha_1$ should reflect the very different ratios of initial surface tension to dynamic pressures as the Knudsen number Kn = $\lambda_a/r_d$ varies among liquids \citep{Ghabache2016,Seon2017} under laboratory conditions, where $\lambda_a$ is the molecular mean free path of air at average ocean atmospheric conditions.

\begin{figure*}[!t]%
\centering
\includegraphics[width=0.95\textwidth]{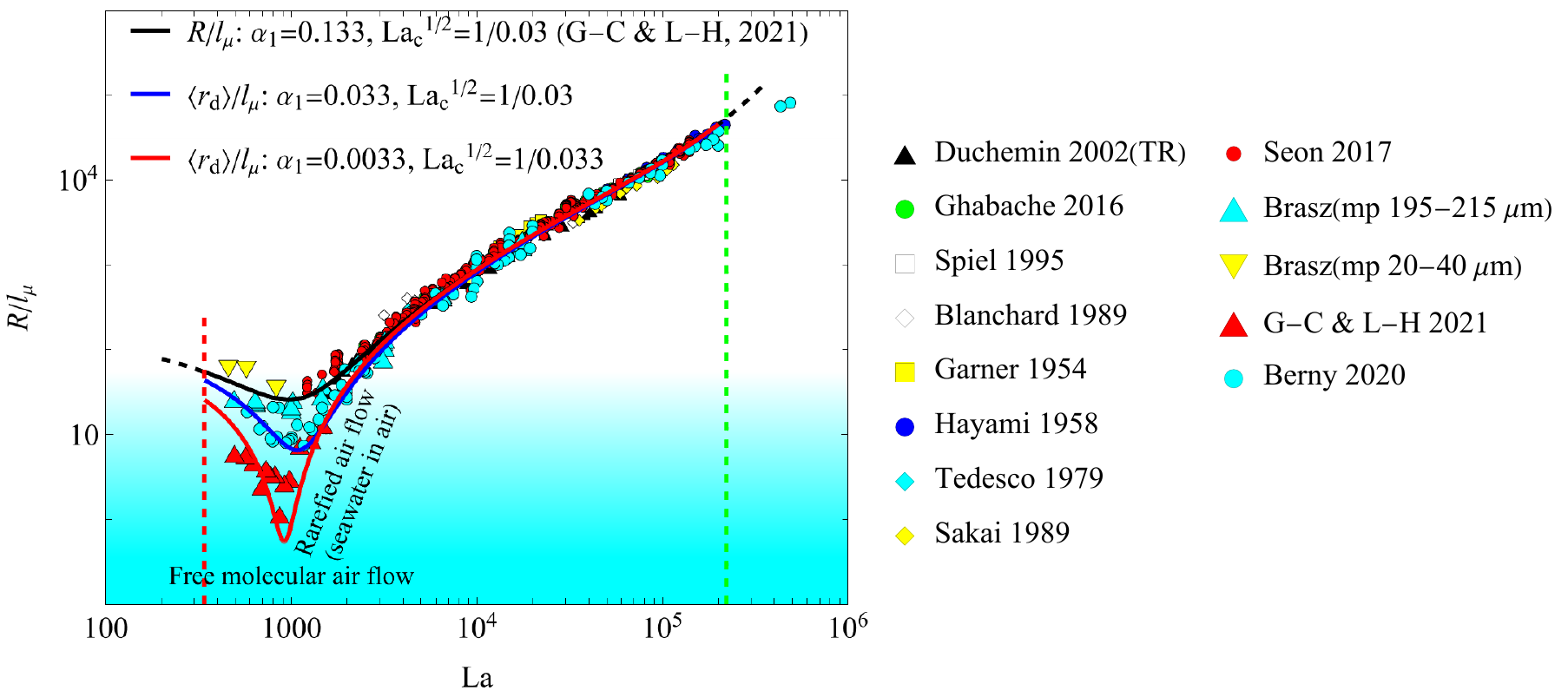}
\caption{Radius of the first ejected droplet $R$ as function of the bubble radius $R_o$ made dimesionless with the viscous-capillary length $l_\mu$, from experimental measurements and numerical simulations taken from the literature (additional information in \citep{GC2021}).
The different sets of La$_c$ and $\alpha_1$ values (continuous lines) fit to different data sets in the range of Laplace numbers La$=R_o/l_\mu$ (abscissae) from the minimum one La$_{min}\simeq 400$ to about La$\sim 2\times 10^5$. Bubbles from 8 $\mu$m to 2 mm cover the range of La numbers marked by vertical red and green dashed lines for seawater properties at $T=15 ^o$C. The shaded cyan region indicates the drop size range for air bubbles in seawater for which rarefied conditions are reached: the cyan intensity represents approximately the hyperbolic tangent of the logarithm of the Knudsen number Kn = $\lambda_a/r_d$ (i.e. $1-2/(1+\text{Kn}^{-2}))$), where $\lambda_a\simeq 68$ nm is the molecular mean free path of air at average ocean atmospheric conditions.
}
\label{f5}
\end{figure*}

Detailed measurements on the first ejected droplet radius $R$ as a function of the normalized bubble radius $R_o$, written as the Laplace number La $= R_o/l_\mu$, and the gravity parameter Bo $= \rho g R_o^2/\sigma$ are available from several authors for an ample collection of experimental and numerical BB measurements \citep{Garner1954,Hayami1958,Tedesco1979,Blanchard1989,Sakai1989,Spiel1995,Duchemin2002,Ghabache2016,Seon2017,Brasz2018,Berny2020,GC2021}. The compilation is shown in figure \ref{f5}, where $R$ is scaled with the natural length $l_\mu=\mu^2/(\rho \sigma)$ and is plotted as a function of La. Continuous lines correspond to the theoretical model \citep{GC2021}:
\begin{equation}
R/l_\mu = k_R \text{La}_c\left(\left(\left(\frac{\text{La}}{\text{La}_c}\right)^{1/2}-1\right)^2
+\alpha_1 \left(\frac{\text{La}}{\text{La}_c}\right)^{1/2}+\alpha_2 \text{Mo}\frac{\text{La}}{\text{La}_c}\right),
\label{RRo}
\end{equation}
with the Morton number Mo $=\frac{g \mu^4}{\rho \sigma^3}$ = Bo La$^{-2}$. The best fitting to available experiments yields $k_R=0.18$, with $\alpha_1\simeq 0.13$ and $\alpha_2 \simeq 0.19$ (\cite{GC2021}, black continuous line).

The different initial conditions of the bursting process and the numerical precision used in simulations may produce a significant variability around a critical La number, La$_c \simeq 1100$  (corresponding to a critical Ohnesorge number Oh$_c\simeq 0.03$, \cite{GC2021}). Interestingly, the data series from each source can be independently and accurately fitted by different $\alpha_1$ values, keeping the same $k_R$ and $\alpha_2$. This fitting parameter $\alpha_1$, which measures the relative magnitude of the surface tension pressure to produce the initial droplet, compared to the dynamic pressure, plays a determining role when La $\simeq$ La$_c$ (see expression \ref{RRo}). Note that the data in figure \ref{f5} correspond to the radius of the {\sl first} ejected droplet, $R$.

However, the emission lasts during times comparable to, or longer than the capillary time $t_o=\left(\rho R_o^3/\sigma\right)^{1/2}$, much larger than the time of formation of the first droplet at the front of the issuing liquid ligament, $t_c=\left(\rho R^3/\sigma\right)^{1/2}$ since $R\ll R_o$. Thus, the high speed liquid ligament has a long time to elongate and disintegrate into a large number of droplets with a variety of radii $r_d$ \citep{Berny2022}. In these conditions, the density and viscosity of the outer medium can dramatically alter the breakup of the spout into droplets.

\subsubsection{High-speed nanometric jet breakup in a rarefied environment}

\begin{figure}[!t]%
\centering
\includegraphics[width=0.75\textwidth]{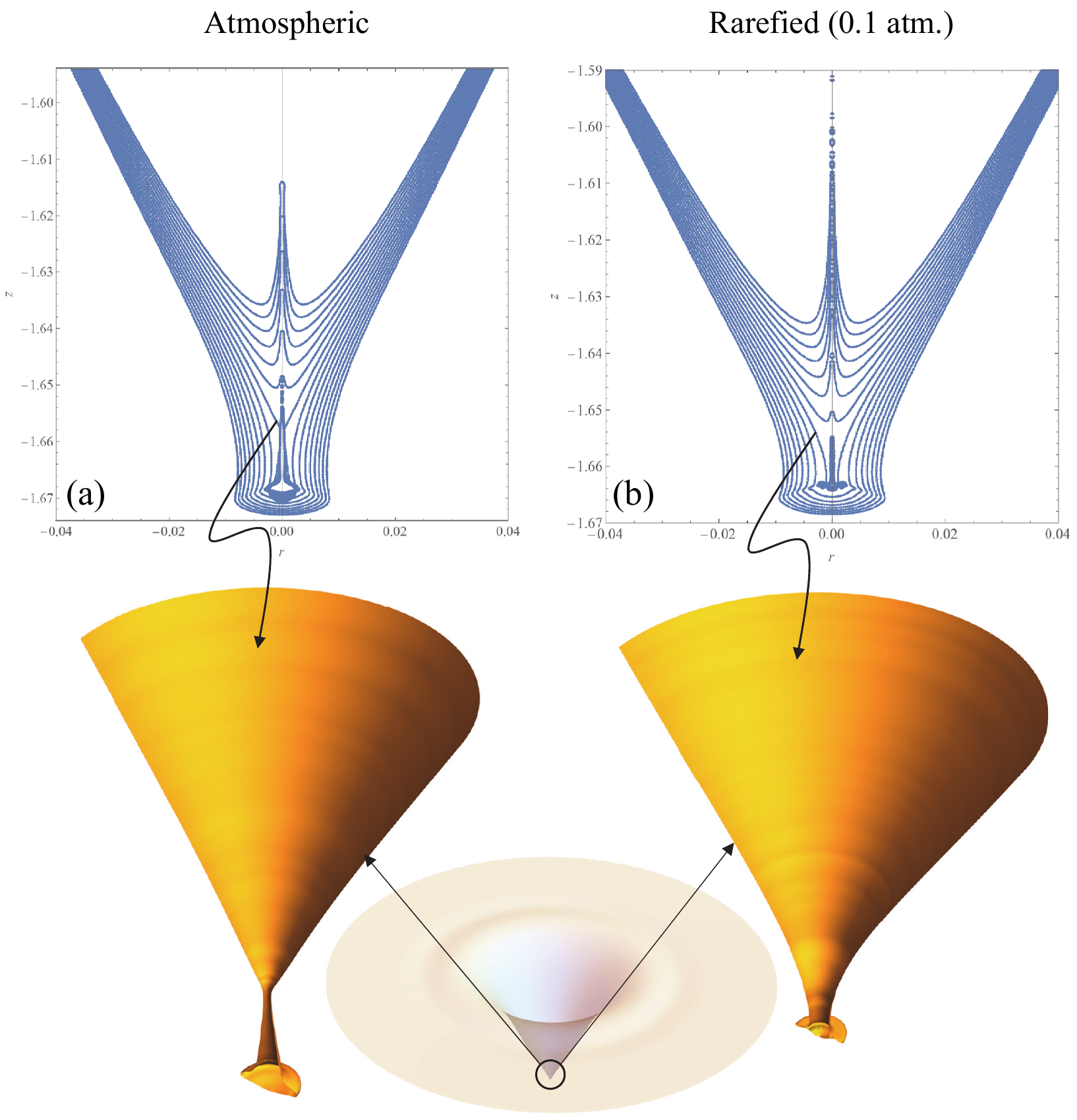}
\caption{The shape of the axisymmetric collapsing free surface at the bottom of the bursting bubble, at 15 successive instants around the time of collapse $t_o$ with constant time intervals $\Delta t=10^{-5} t_c=1.9245 \times 10^{-3} t_\mu$, for Oh = 0.03. Density and viscosity ratios  $\varphi=\rho_g/\rho=0.001$ and $\eta=\mu_g/\mu=0.01$, respectively, and (b) $\varphi=0.0001$, $\eta=0.001$ (rarefied gas conditions, 0.1 atmospheres). Numerical simulations from \citep{GC2021} made with Basilisk \citep{Basilisk} using level 16. Observe the shape of the collapse neck: it is nearly at the bottom in the rarefied conditions.}
\label{f6}
\end{figure}

The impact of the environment rarefication was demonstrated reducing the viscosity and density of the outer environment one order of magnitude (see figure \ref{f6} in \cite{GC2021}), which keeps the high velocity of the jet front for a longer time. Indeed, the bubble radius corresponding to La$_c$ (minimum $r_d$) is $R_o\simeq 20\,\mu$m. This would lead to drop radii well below the mean free path of gas molecules of the environment, and ejection speeds above their average molecular speed. Hence, the values of the fitting constant $\alpha_1$ should reflect the very different ratios of initial surface tension to dynamic pressures as the Knudsen number Kn = $\lambda_a/r_d$ varies among liquids \citep{Ghabache2016,Seon2017} under laboratory conditions, where $\lambda_a$ is the molecular mean free path of air at average ocean atmospheric conditions.

A brief inspection of the values attained for the sizes and speeds of these droplets around a critical value La$_c\simeq 1100$ (e.g. \cite{Seon2017,Berny2020,GC2021}) for seawater shows that they are indeed in the range of ultra-fine aerosols, with sizes well below the molecular mean free path (around 70 nm in air at standard conditions) and velocities exceeding by far the average molecular speed of the surrounding gas (around 290 m/s). In these extreme cases, the consideration of a stochastic, extremely rarefied environment would be the most realistic assumption. No one has ever simulated this complex phenomenon, let alone directly observed it, completely beyond the capabilities of current measurement techniques, and its direct visualization or assessment is impossible.

The evolution of high speed nanometric jets in vacuum using molecular simulations has been reported in the literature \citep{Moseler2000}. These simulations show the rapid action of surface tension, even under a non-continuous approach, in terms of equivalent local Weber and Ohnesorge numbers We = $\rho v^2 d_j/\sigma$ and Oh = $\mu/(\rho \sigma d_j)^{1/2}$, respectively, where $d_j$ (much smaller than $l_\mu$) is the local diameter of the liquid ligament. To achieve a high ejection velocity bypassing the action of viscous forces at these extremely small scales, a huge energy density much larger than $\mu^2/(\rho d_j^2)$ should be locally applied. For seawater and $d_j$ around 5 nm, the energy density involved should be greater than about $5\times 10^7$ Pa.

In this regard, observe that the collapse of the neck occurs much closer to the bottom (i.e., the gas volume of the trapped bubble is significantly reduced) under rarefied conditions, which produces an enhanced kinetic energy focusing at the instant of collapse and a significant excess of ejection velocity compared to the atmospheric conditions. This is a key consideration that also applies to other similar jetting processes like flow focusing or electrospray \citep{Montanero2020} that leads to the formation of nanometer-sized droplets \citep{Rosell1994}. Once the jet is ballistically ejected, the local action of surface tension immediately promotes the fragmentation of the ligament and the production of droplets \citep{Villermaux2004} if the dynamical effect of the environment is negligible (e.g. rarefied gas or vacuum).

\begin{figure}[!t]%
\centering
\includegraphics[width=0.85\textwidth]{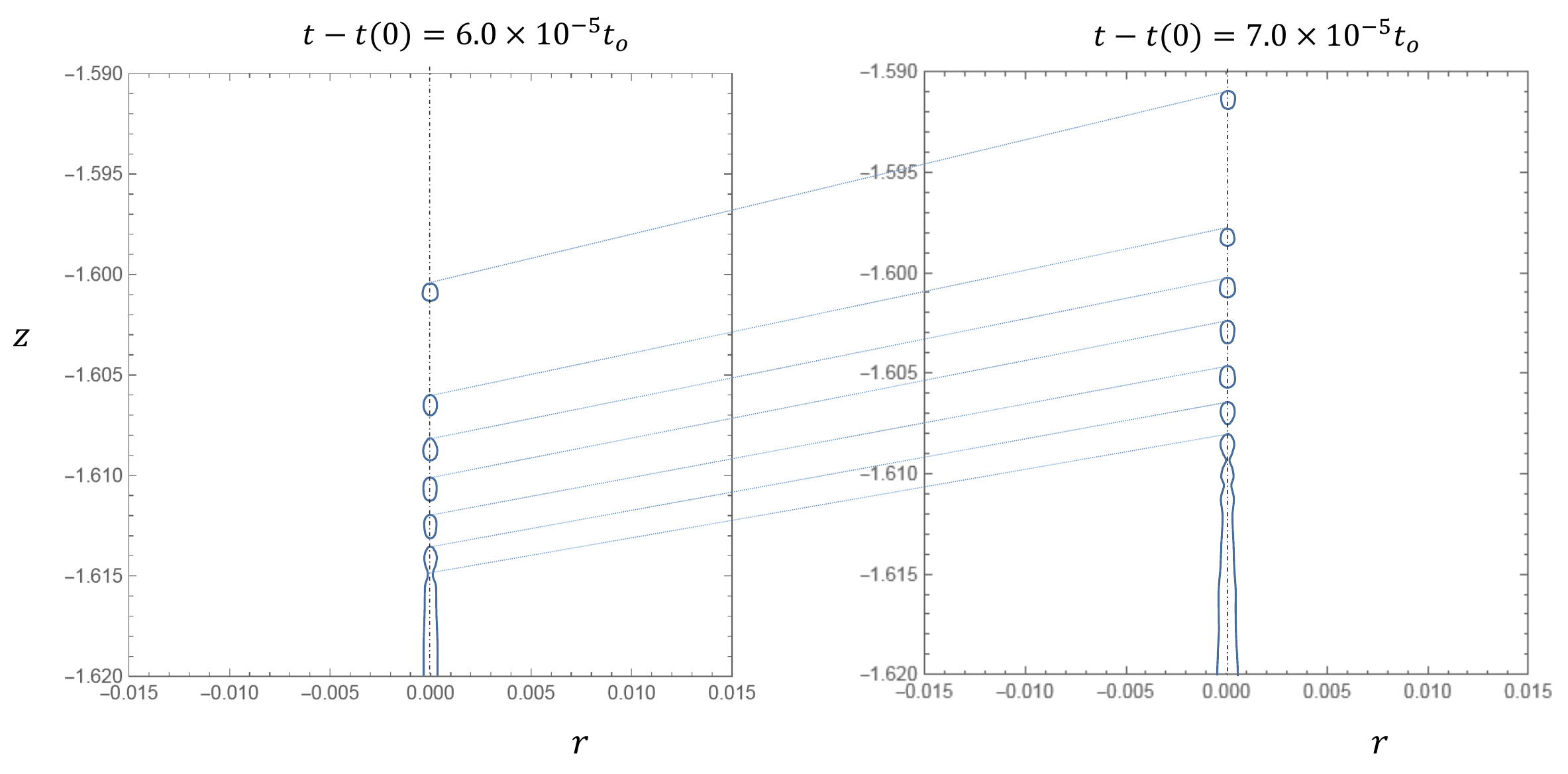}
\caption{Two successive instants of the ballistic ejection of the first six droplets for La = 1111 (Oh = 0.03), and for rarefied conditions compared with those the water and air at atmospheric conditions at average sea level and temperature ($\varphi=0.0001$, $\eta=0.001$) \citep{GC2021}. The resulting radii are about 7.5 nm for seawater.}
\label{f7}
\end{figure}

In contrast, the numerical simulations of BB made so far \citep{Duchemin2002,Berny2020,GC2021} assume the continuum hypothesis \citep{Basilisk}, which excludes the possibility of a rarefied or vacuum environment and the early action of surface tension observed in real conditions. However, even under continuum assumptions, a reduction of the outer density fosters the early spout breakup, too. An illustration of the onset of an extreme emission event (very small sizes and large ejection velocities) with a reduced density and viscosity gas-liquid ratios ($\varphi=\rho_g/\rho$ and $\eta=\mu_g/\mu$, respectively) is given in figure \ref{f7}. It shows a sequence of up to 6 extremely small initial droplets ejected consistently with previous statistical analysis \citep{Villermaux2004,Berny2021} if the number of ejected droplets $n_d$ is sufficiently large. Indeed, from the values of the two subsequent times here considered, note that the initial frequency of droplet ejection is extremely high, about $4\times 10^5\, t_o^{-1}$. 

Thus, the central spout ejected by small collapsing bubbles in seawater could yield much more numerous droplets, their size could be much smaller, and their ballistic speed much larger than previously expected. Numerical simulations show \citep{GC2021} that their size, production frequency and speed can reach values well beyond the natural scales of seawater at the average temperature of the ocean (15$^o$C), i.e. $l_\mu=\mu/(\rho \sigma) = 19.5$ nm, $t_\mu^{-1} = \mu^3/(\rho \sigma^2) = 0.32$ ns, and $v_\mu$, respectively. The size, frequency and velocity of the ejected droplets in simulations can be, respectively, 10 to 20 nm (depending on temperature and salinity), around the GHz frequency, and 300 to 600 m/s. (\cite{GC2021}, i.e. beyond the thermal speed of molecules of the surrounding gas air). Consequently, they can reach distances well beyond any previous considerations.

\subsubsection{Earlier evidences of the role of ultrafine jet drops}

Regarding the chemical composition of the measured spray, \cite{Wang2017} made a fundamental insight to determine differences that could be ultimately assigned to either film or jet drops. In fact, the film droplets are richer in species of the molecular layers closer to the surface. In contrast, despite the presence of a recirculating region (see figure \ref{f4}), the streamline pattern indicates that the material ejected as jet droplets should be a sample from the liquid bulk \citep{GC2021}, not the surface, consistently with the results from \cite{Wang2017}. The results of Wang et al. raised a crucial issue in the field. However, as the bubble size decreases down to the micrometric scale, the liquid sample in the droplets should be increasingly dominated by the sea surface microlayer composition \citep{Cunliffe2013}, even for jet droplets. In addition, given the extremely small size of the liquid relics, their acidity and consequent reactivity could reach high levels \citep{Angle2021}, together with their capability to immediately nucleate or react with volatile organic components (VOC) present in the surrounding atmosphere \citep{Mayer2020}. Thus, below certain submicron droplet size, to observe distinctions of kinematic origin from the physicochemical nature of the eventual seawater aerosol becomes impossible with current equipment and experimental setups. Nonetheless, the findings by \cite{Wang2017}, the analyses of \cite{Berny2020,Berny2021,Berny2022}, and previous considerations would point to the jet droplets as a potential origin of at least a major fraction of submicron SSA and SMA present in the atmosphere.

Interestingly, \cite{Wang2017} specified $R_o=13$ $\mu$m and 4 $\mu$m as the bubble size responsible of the highest measured sound frequency from breaking waves \citep{Dahl1995,Deane2010} and the minimum bubble size capable of producing jet drops \citep{Lee2011}, respectively, without resorting to any physical description of the bubble breakup mechanism. In reality, the two bubble sizes $R_o=13$ $\mu$m and 4 $\mu$m aforementioned approximately correspond to the two key values of the Laplace numbers La$_c=1111$ and La$_{min}=343$, respectively, reported in the literature (i.e. Oh$_c=0.03$ and Oh$_{min}=0.054$, respectively, \cite{Duchemin2002,Ghabache2016,Seon2017,Walls2015,GC2017,Berny2020,GC2021}). These Laplace numbers correspond to:

(i) The minimum radius of the first ejected droplet for the whole spectrum of jet emitting bubbles, and

(ii) The minimum bubble radius $R_{o,min}$ for which jet droplets are ejected.

\subsubsection{The probability density function of jet droplet size}

Given the difficulty of direct measurements of jet droplet statistics, researchers have relied on numerical simulation. Deike and coauthors \citep{Berny2021,Berny2022} studied the variability of data sizes along the whole transient jet emission event, an essential ingredient of any statistical global model. These authors offer a very ample set of data under ``noisy initial conditions'' for all-drops ejected droplet radii $r_d$ from simulations, with La numbers spanning close to three orders of magnitude.
A best fitting statistics of these data to a Gamma distribution \citep{Villermaux2004}, as
\begin{equation}
    p(\chi_{d,i})=\frac{\alpha^\alpha}{\Gamma(\alpha)} \chi_{d,i}^{\alpha-1}\exp(-\alpha \chi_{d,i}),
\end{equation}
yields a shape factor $\alpha$ dependent with La approximately as $\alpha=0.65$ La$^{1/5}$. Here, $\chi_{d,i}=r_{d,i}/\langle r_d \rangle$. Note that $r_d$ is, for each bursting event, a statistical variable where its average $\langle r_d \rangle$ is a function of La with the same form as (\ref{RRo}) but a different $\alpha_1$ value (i.e. a different ratio of surface tension to kinetic energy at the front of the spout) than the one of the first droplet $R$: in general, $R\neq \langle r_d \rangle$. Note the appearance of the power $1/5$ affecting La in the jet droplet like in the film flapping droplets statistics, whose analysis is beyond present study.

\begin{figure}[!t]%
\centering
\includegraphics[width=0.750\textwidth]{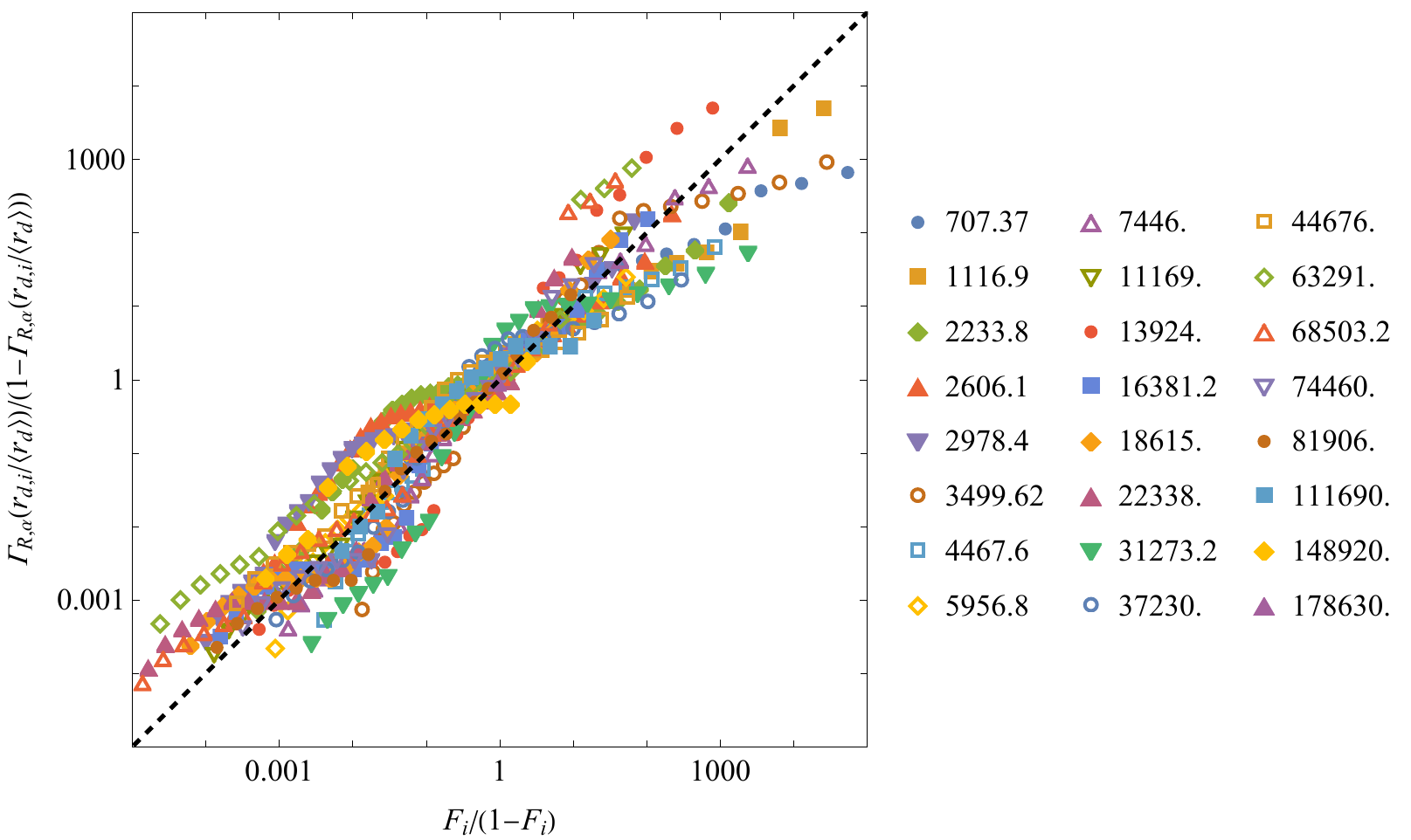}
\caption{Test of the data from Berny et al. \citep{Berny2021} against a Gamma function. The test follows a modified Anderson-Darling approach, see text. The right labels are the La numbers of each data series.}
\label{f8}
\end{figure}

The fitting used a modified Anderson-Darling test as follows. The data are represented in figure \ref{f8} as the cumulative distribution value $F_i$ for each $i-$droplet of radius $r_{d,i}$ divided by $(1-F_i)$, versus the theoretical values corresponding to a Gamma distribution, i.e. $\Gamma_{R,\alpha}(\chi_{d,i})/\left(1-\Gamma_{R,\alpha}(\chi_{d,i})\right)$, where $\Gamma_{R,\alpha}(\chi_d)=\Gamma(\alpha,0,\alpha \chi_d)/\Gamma(\alpha)$ is the normalized cumulative Gamma distribution function. Observe the reasonable global statistical goodness-of-fit for the whole range of La numbers explored (about three orders of magnitude). This guarantees a sufficient confidence on the assumption that the breakup of the time-evolving ejected liquid ligament approximately follows a Gamma distribution as predicted by \cite{Villermaux2004}, with a La-dependent shape factor $\alpha=0.65$ La$^{1/5}$.

\subsection{The number of ejected droplets per bursting event}\label{sec5}

\subsubsection{Film drops}

\cite{Jiang2022} give very useful experimental values of the number of droplets ejected per bursting event in their figures 3B and S6. Their collection of data from other authors is especially useful too. These data are gathered in figure \ref{f10}, where the two mechanisms, film bursting and flapping, are separately considered according to their characteristic number production.

A particularly notable finding of \cite{Jiang2022} is their explanation of the Blanchard-Syzdek paradox \citep{Blanchard1988} around La$_{BS}\simeq 7 \times 10^4$ for seawater ($R_o\simeq 1.4$ for $T=15^o$C to 2.3 mm for $T=5^o$C), where the film flapping mechanism dominates for La $<$ La$_{BS}$ while the film bursting does so for La $>$ La$_{BS}$. Since Jiang et al. consider $R_o$ as twice that of the equivalent spherical volume radius, their own data do not appear to collapse well with those of the other authors. However, taking $R_o$ as the equivalent spherical radius assumed by the other authors, the collapse of Jiang's data with the rest of authors is evident.

Despite one observes certain degree of overlapping between the film breakup mechanisms due to the complexity of the collective bursting process and its critical dependency on different factors (temperature, presence of surfactants, etc.), there is a relatively clear statistical separation between them. For the purposes of global modelling, one could separately fit the number of droplets produced by each mechanism by the expressions:
\begin{equation}
n_{d,B}=N_B \frac{\text{La}}{\varpi \text{La}_{BS}} \exp\left(-\frac{\varpi \text{La}_{BS}}{\text{La}}\right)
\label{ndB}
\end{equation}
for the film bursting droplets, with $N_B\simeq 40$, and:
\begin{equation}
n_{d,F}=N_F \left(\frac{\text{La}}{\text{La}_{BS}}\right)^{\alpha_F}\left(1-\frac{\text{La}}{\text{La}_{BS}}\right),\quad (\text{valid for La} < \text{La}_{BS})
\end{equation}
for the film flapping droplets, with $N_F\simeq 600$ and $\alpha_F \simeq 2.75$. The overlapping is reflected by the factor $\varpi\simeq 1.57$ affecting the value of La$_{BS}$ in (\ref{ndB}). These fittings are plotted in figure \ref{f10}.

The number of film bursting droplets tend to scale with La for La$\rightarrow \infty$ since their number would be proportional to the radius of the equivalent volume sphere due to mass conservation: the fragmenting film rim has a length commensurate with the bubble radius while the film thickness becomes nearly independent of La. However, the marginal dependence of the film flapping droplet number $n_{d,F}$ for decreasing La with the power $\alpha_F$, between the bubble surface ($\alpha_F\sim 2$) and volume ($\alpha_F\sim 3$), could be related to the decreasing effect of the gas in the bubble and its surroundings by rarefaction (i.e. when Kn increases) as the bubble size and consequently the submicron droplet sizes decrease. The fundamental effect of the bubble gas is well documented by \cite{Jiang2022}. On the other hand, flapping droplets seem to cease abruptly at the Blanchard-Syzdek transition value La$_{BS}$.

\subsubsection{Jet drops}

The number of ejected droplets is one of the main claims of this work: for seawater, this number can be orders of magnitude larger than any previous estimation based on experiments with other liquids in air or numerical simulation assuming a continuum gas atmosphere. In the bubble size range from about 10 to 100 micrometers, seawater produces ejections with associated Knudsen numbers above unity and velocities beyond the thermal speed of air. These two facts would lead to much larger droplet fragmentation frequencies (and total ejected droplet number) than previously thought.

The main cause of the extremely large velocity of the issued liquid spout is the radial collapse of an axisymetric capillary wave at the axis \citep{Walls2015,GC2018,GC2021}, producing a singularity \citep{Eggers2007} and the trapping of a tiny bubble at the bottom of the cavity (see figure \ref{f6}). The radial collapse elicits a subsequent ejection in the axial direction of an initially quasi-infinitesimal spout of liquid at an extremely large velocity: observe that the initial droplet ejection velocity in figure \ref{f7} can be about $10^3\times V_o$, where $V_o=\left(\sigma/(\rho R_o)\right)^{1/2}$ is the characteristic velocity of the bursting process driven by capillary forces.

\begin{figure}[!t]%
\centering
\includegraphics[width=0.65\textwidth]{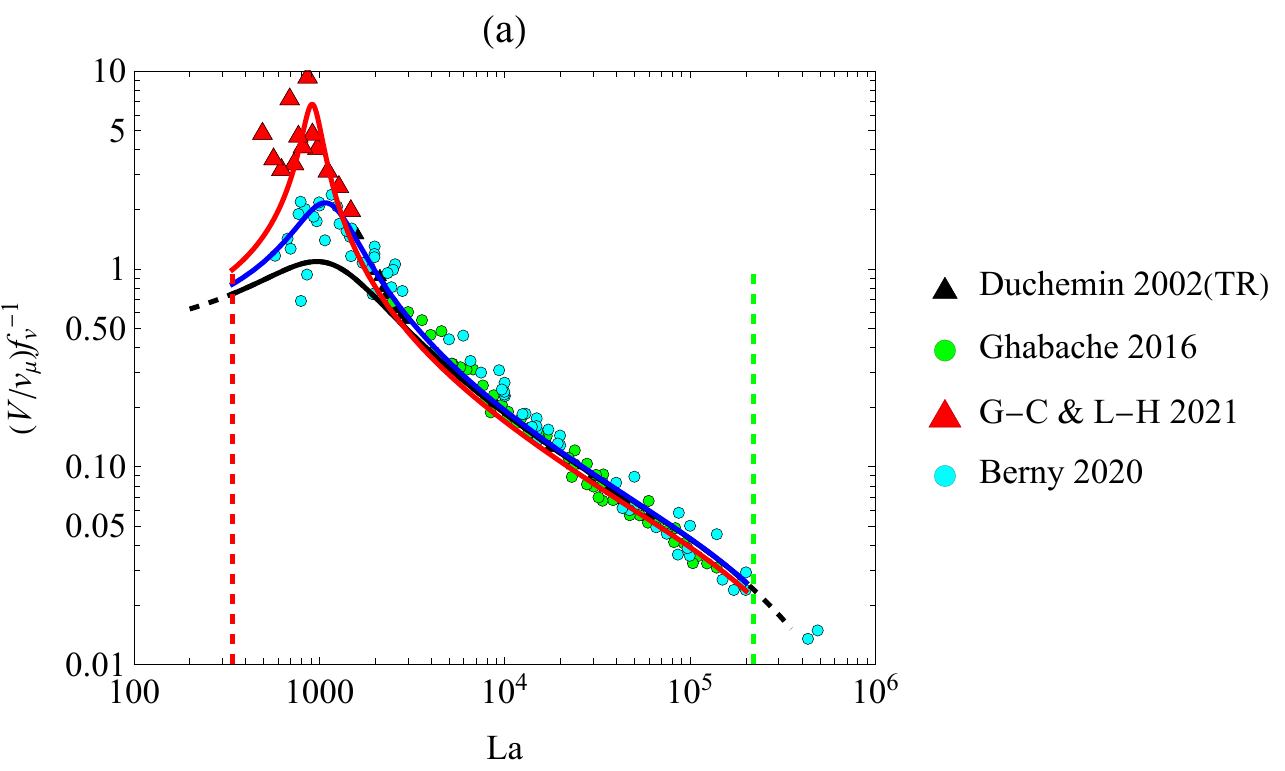}
\caption{The measured ejection velocity $V$ of the first drop, corrected with the factor $f_v$ and compared with the proposed model. The curves correspond to the velocity of the first droplet (black line), as fitted in \cite{GC2021}, and the alternative fitting here proposed for the average $\langle v_d\rangle$ (blue line).}
\label{f9}
\end{figure}

The ejection velocity of the first droplet $V$ has been investigated in \cite{Duchemin2002,Ghabache2016,Seon2017,GC2017,GC2018,Berny2020,GC2021}, among other works. Fitting the speed of emission is trickier than the droplet size, given the dependency of the former on the point at which it is measured and the inherent variability of the liquid spout velocity. If that point is set approximately at the point where the droplet is released, the recent model proposed by \cite{GC2021} yields:
\begin{equation}
V/v_\mu= k_V \left(\left(\left(\frac{\text{La}}{\text{La}_c}\right)^{1/2}-1\right)^2
+ \alpha_1 \left(\frac{\text{La}}{\text{La}_c}\right)^{1/2}+\alpha_2 \text{Mo}\frac{\text{La}}{\text{La}_c}\right)^{-1/2},
\label{VVo}
\end{equation}
This model is compared with the experimentally measured $V$ in figure \ref{f9}. In general, the scale of the ejection velocity $v(t)$ co-evolves with the radial scale of the ejection $r(t)$ as $r(t) \sim v(t)^{-2}$ along the process (see \cite{GC2021}). This trend is consistent with the valuable data provided by \cite{Berny2020} for the five first ejected droplets (see their figures 6 and 7). However, while the prefactor $k_R$ is a constant, the best fitting to the experimentally measured and reported $V$ demands that $k_V$ (of order unity) should be slightly dependent on La (Oh) and Bo as $k_V=0.39 f_v($La,Bo$)$. The fitting function $f_v$ proposed by \cite{GC2021} was $f_v=\left(1+ k_1 \text{Bo} +k_2 \text{Oh}\right)^{-1}$ with $k_1=2.27$ and $k_2=-16$. A better fitting is here obtained with $f_v=\left(1+ k_1 \text{Bo} + \text{La}^{\gamma_1} \right)^{-1}\left(1- (\text{La}_{min}/\text{La})^{\gamma_2} \right)$, with $\gamma_1=-0.125$ and $\gamma_2=0.8$

The extremely large initial velocity of the incipient spout rapidly decays as the mean radius of the ejected spout increases along the ejection process. According to high precision numerical simulations \citep{Deike2018,GC2021} assuming seawater and air at atmospheric conditions, it decays about an order of magnitude before the first drop is released. Interestingly, the speed of the first droplet $V$ is comparable to or larger than the capillary-viscous or natural velocity $v_\mu=\sigma/\mu$ for La in the range from La$_{min}\simeq 400$ to about $4\times 10^3$ (see figure \ref{f9}(a)), the range where the maximum ejection velocities and minimum jet droplet radius are reached, and where the bottom micro-bubble trapping is prevalent.

For the critical La$_c=1100$ values where $V$ peaks (where $f_v$ is about 0.52), $V$ can reach values as high as one order of magnitude above $v_\mu$ \citep{GC2021}. In contrast to the rest of the domain, which shows robustness to initial perturbations \citep{Berny2022}, the parametrical region around La$_c$ is highly sensitive to effects like the gas conditions, as previously explained. In \cite{GC2021} (see figure 2 in that work) we showed that the transversal size of the ejected spout is inversely proportional to its velocity for times smaller than the natural one $t_\mu$ from the instant of collapse. Hence, the sooner the ballistic jet starts ejecting droplets, the smaller, faster and more numerous those droplets will be. In that initial time interval, the velocity of ejection in the simulations (assuming incompressibility) can be as high as 20 times the natural velocity $v_\mu$ (about 61 m/s for seawater), with spout sizes about 0.1 to 0.2 times $l_\mu\simeq 19.5$ nm. In the absence of interactions with the environment, or in conditions of minimized interactions, that initial velocity can overcome the speed of sound in seawater, and -consistently- the emitted droplets can be orders of magnitude smaller than the molecular mean free path of air in standard conditions.

According to \cite{Chandra1961} and in the absence of any interaction with the environment, the most unstable wavelength $\lambda$ for a viscous liquid column of radius $r_j$ is equal to $\lambda=2 \pi \psi r_j/k$, with $k\simeq 0.697$ and $\psi$ a function of Oh$_j=\mu/(\rho \sigma r_j)^{1/2}$ very approximately equal to $\psi=(1+2 \text{Oh}_j)^{1/2}$. Given that the environment surrounding the ejected ligament is a high speed gas jet (see figure \ref{f4}) co-flowing alongside with that, one can assume that the liquid is moving with a comparable velocity to that of the environment and its dynamical effect can be, in first approximation, neglected. Thus, the conservation of mass on breakup leads to:
\begin{equation}
    r_d/r_j\equiv \zeta=\left(\frac{3 \pi}{2\times 0.697}\right)^{1/3}\left(1+ 2(\zeta /\chi_d )^{1/2}\right)^{1/6}
    \label{rj}
\end{equation}
where $\chi_d=r_d/l_\mu$. This is a transcendental function for $\zeta$ whose solution can be very approximately resolved for a given $\chi_d$ using the fixed point method with initial value $\zeta=1$ in the right hand side of (\ref{rj}). A couple of iterations (that can be explicitly expressed) yield the solution with maximum errors below 0.1\%. The relationship (\ref{rj}) is expected to hold even close to the molecular scale, as demonstrated by \cite{Moseler2000,Zhao2020} via molecular simulations of nanojets.

Next, the instantaneous ejection flow rate is proportional to $v_d r_j^2$. Obviously, there is an inherent variability of the ejected droplet radius and velocity along a single bursting, implying the consideration of the two stochastic variables $r_d$ and $v_d$ in the calculations. However, one has that:

(1) There is a strict limitation for the time along which the ejection is active, given by $t_o=\left(\rho R_o^3/\sigma\right)^{1/2}$,

(2) As previously seen, both stochastic variables $\chi_d$ and $\upsilon_d=v/v_\mu$ have well defined statistics in a single bursting event, with average $\langle \chi_d\rangle$ and $\langle \upsilon_d\rangle$ depending on La and Mo only. These average values should be (universally) proportional to both $R/l_\mu$ and $V/v_\mu$ in a single bursting event.

Thus, the conservation of total mass along a single ejection event within a continuous bursting of bubbles of different sizes demands:
\begin{equation}
    n_d \sim t_o \frac{v_d}{r_d}\left(\frac{r_j}{r_d}\right)^{2} \sim t_o \frac{\langle v_d\rangle}{\langle r_d\rangle} \zeta^2 \Longrightarrow n_d=k_c \text{La}^{3/2} \frac{\langle \upsilon_d\rangle}{\langle \chi_d\rangle} \zeta^2,
    \label{nd}
\end{equation}
where both $\langle \chi_d\rangle$ and $\langle \upsilon_d\rangle$ are given by the expressions (\ref{RRo}) and (\ref{VVo}), with fitting constants $k_{R,V}$ and $\alpha_1$ that can be different from the original ones for $R$ and $V$, summarized by the constant $k_c$. This is justified since
(i) $\alpha_1$ measures the relative weight of surface tension energy to form the droplet at the front of the issuing liquid spout \citep{GC2021}, which can vary along the bursting event, and
(ii) the prefactors $k_{R,V}$ should obviously reflect the overall change of both $r_d$ and $v_d$ along the bursting, too.

Hence, one can calculate the single-event averaged number of ejected droplets using (\ref{nd}) as a function of La and Mo alone. Note that Berny et al. \citep{Berny2021} sought for a scaling law as $\langle n_d\rangle \sim$ La$^{-1/3}$ in their recent work. They directly measured the size of ejected droplets using numerical simulations with fixed density and viscosity ratios with the environment. However, our simulations point to the appearance of an enormously larger number of ejected droplets as one decreases those ratios (see figure \ref{f7}) or when the gas environment becomes rarefied close or beyond the molecular mean free path scale, which may apply to the case of water in air for La around La$_c$.

\begin{figure*}[!t]%
\centering
\includegraphics[width=0.85\textwidth]{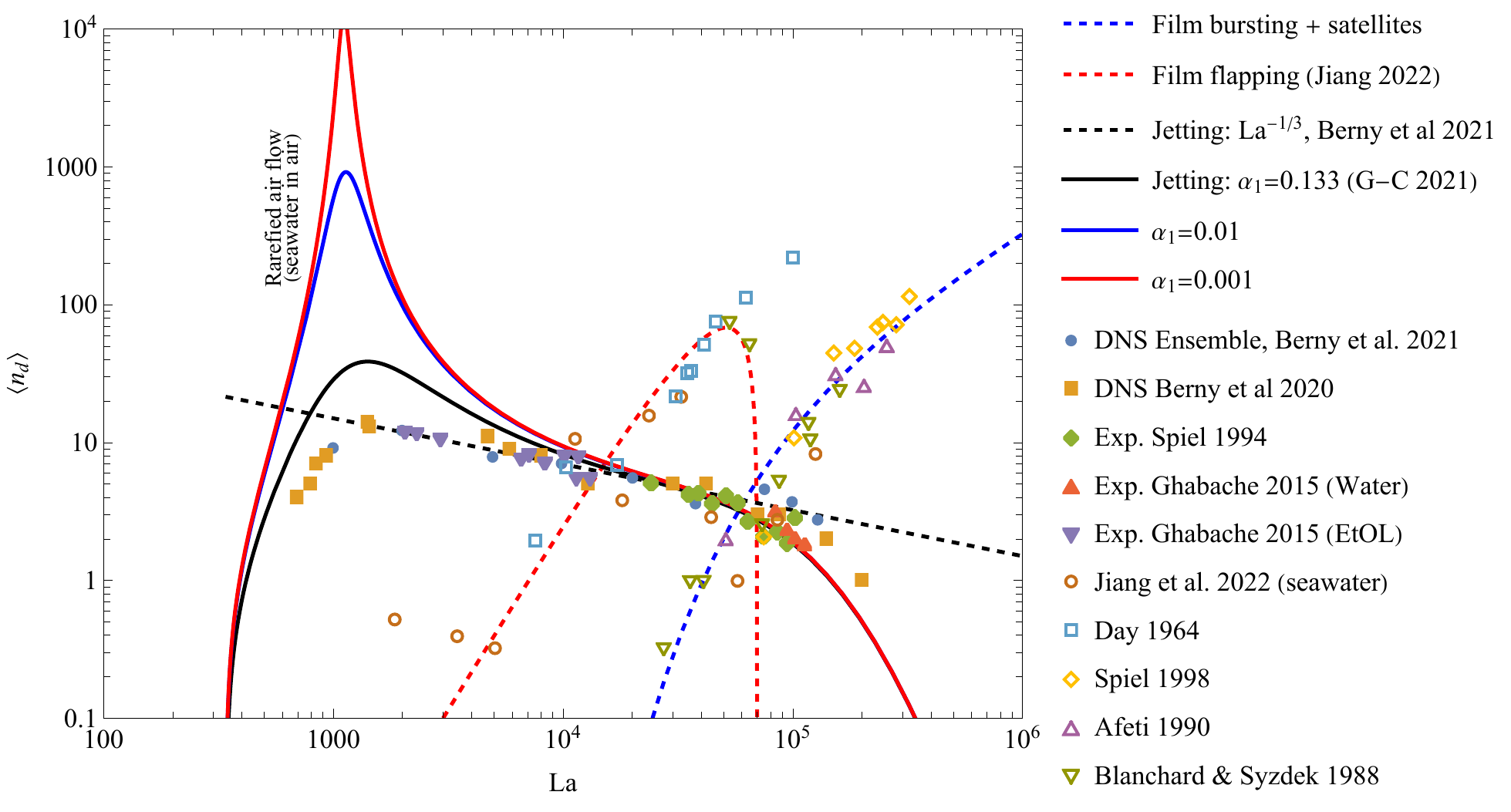}
\caption{The average number of droplets ejected per bursting event, $\langle n_d \rangle$. Data gathered from \citep{Berny2021}. The continuous lines correspond to the theoretical model (\ref{nd}), for different $\alpha_1$ values and Mo calculated for seawater at 15$^o$C. Observe that the size range of bubbles effectively shooting jet droplets is from about 6-7 micrometers (La $\simeq 350$) to about 2 mm (La $\simeq 10^5$). A direct measurement, visualization or simulation of ejections for seawater around La$_c$ is impossible: only an indirect assessment is viable.}
\label{f10}
\end{figure*}

Indeed, observe that the fitting constant $\alpha_1$ in (\ref{RRo}) and (\ref{VVo}) strongly determines both the minimum droplet radius and the maximum ejection speed for the whole La-span, and therefore the total number of droplets ejected in figure \ref{f10}. As previously noted, the numerical simulations show that the bursting in a rarefied environment conditions can lead to extraordinarily small droplet radii and enormous ejection speeds, resulting in a surprisingly large number of ejected droplets around the critical La$_c$ (see figure \ref{f10}).

Fitting the number of droplets ejected per bursting event $n_d$ for seawater to the data set gathered by \cite{Berny2021} for La $> 10^4$ approximately yields $k_c$ = 1. However, the average number of droplets generated could deviate very significantly from Berny's assumption (i.e. $n_d \sim $ La$^{-1/3}$) for La $<10^4$ assuming rarefied gas conditions, which entails using different $\alpha_1$ values. Note that the new approach correctly predicts no ejection (i.e. $\langle n_d\rangle\rightarrow 0$) as La approaches the two extreme values already mentioned in the literature \citep{Berny2020,Berny2021}.

\subsection{The bubble statistics}\label{sec7}

\begin{figure}[!t]%
\centering
\includegraphics[width=0.85\textwidth]{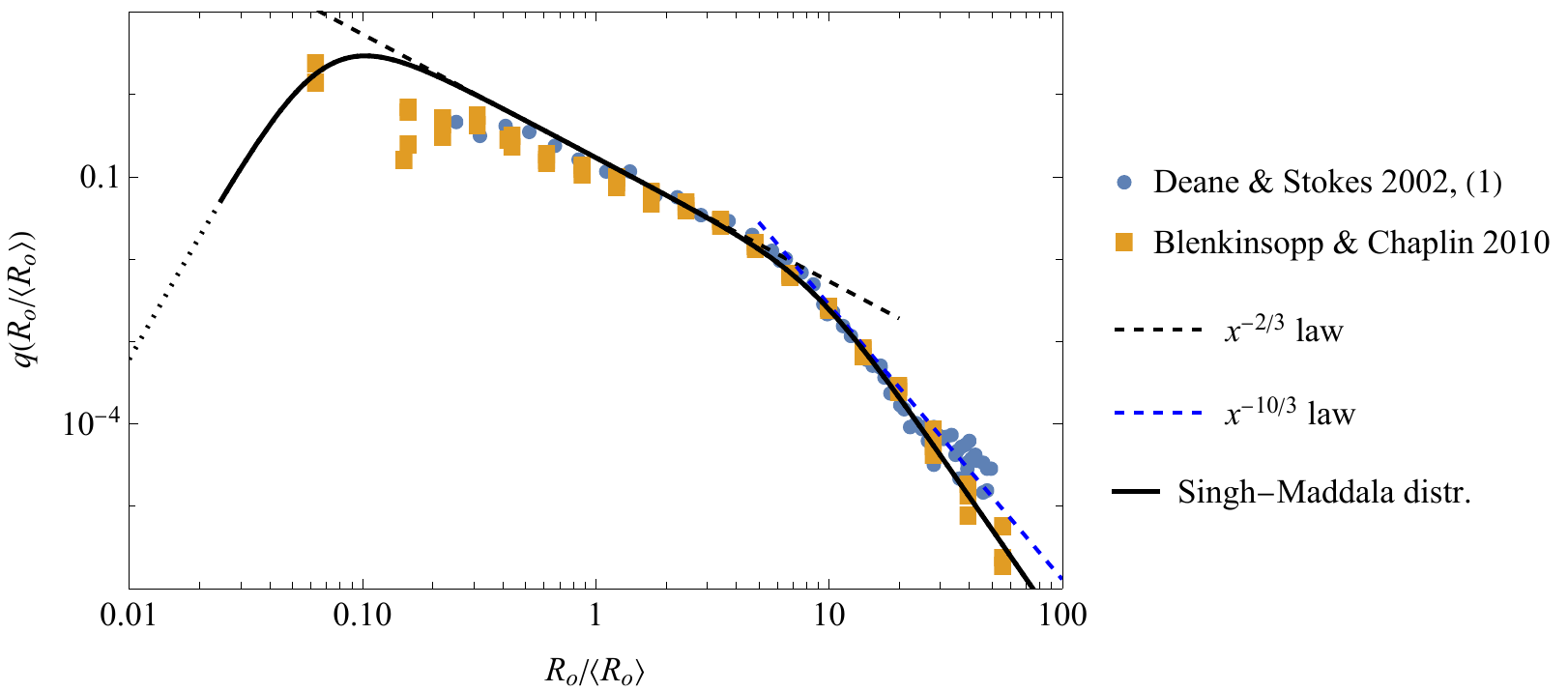}
\caption{Probability distribution of the bubble radius $R_o$ produced by breaking waves, made dimensionless with the average bubble radius $\langle R_o\rangle$ of the best fitting continuous probability distribution to the data sets from \citep{Deane2002} and \citep{Blenkinsopp2010} (see text).}
\label{f11}
\end{figure}

There is an ample literature on the subject of bubble generation in the ocean and its qualitative analysis (e.g. \cite{Cipriano1981,Dahl1995,Deane2002,Blenkinsopp2010,Al-Lashi2018}). The data from \cite{Deane2002} and \cite{Blenkinsopp2010} have been established as a reliable source of experimental measurements of bubble plumes and swarms produced by breaking waves. Figure \ref{f11} plots both data sets, where both the sub- ($\sim x^{-3/2}$) and super-Hinze ($\sim x^{-10/3}$) scales \citep{Deane2002} are clearly visible. The bubble radius $R_o$ is made dimensionless with the average bubble radius $\langle R_o\rangle$ obtained from the best fitting continuous probability distribution to the data sets from \cite{Deane2002} and \cite{Blenkinsopp2010}, which yields $\langle R_o\rangle \simeq 0.25$ mm. From the fundamental theoretical considerations on the bubble dynamics made by \cite{Clarke2003,Quinn2015}, showing the existence of two clearly defined power-law ranges and a drastic decay below a certain $R_o$, the use of an analytic extended Singh-Maddala probability density distribution (p.d.f.) $q(x)$ is proposed in this work as follows:
\begin{equation}
 q(x)=A\,a\, x^{a-1}\left(1+\left(\frac{x}{\epsilon X}\right)^d \right)^{\frac{1-a+b}{d}} \left(1+\left(\frac{x}{X}\right)^d\right)^{\frac{-b+c}{d}},
 \label{SMD}
\end{equation}
with $b=-3/2$ \citep{Clarke2003,Quinn2015}. Constants $A$ and $X$ can be analytically obtained from the condition that the $s-$moments distribution, given by:
\begin{equation}
 F(x,s)= A\, x^{a+s} F_1\left(\frac{a+s}{d},\frac{b-c}{d},\frac{-1+a-b}{d},\frac{a+d+s}{d},
 -\left(\frac{x}{X}\right)^d, -\left(\frac{x}{\epsilon X}\right)^d \right),
\end{equation}
are equal to 1 for both $s=0$ and 1, i.e. making $F(x\rightarrow \infty,s=0)=F(x\rightarrow \infty,s=1)=1$. $F_1$ stands for the first of the Appell hypergeometric series. Besides, $x=R_o/\langle R_o\rangle=$ La $\,l_\mu/\langle R_o\rangle$, $X=$ La$_{Hinze} l_\mu/\langle R_o\rangle$, and $\epsilon=$ La$_{min}/$La$_{Hinze}$, where La$_{Hinze}$ is the value of La for the Hinze bubble radius, about 2 mm \citep{Clarke2003}, i.e. La$_{Hinze}\simeq 1.5\times 10^5$.

Alternatively, $q$ can be expressed as a p.d.f. for the stochastic variable La as $q(x)=\langle$La$\rangle \theta(La)$, where the first moment of $\theta($La$)$ ($s=1$) is $\langle $La$\rangle$.
Regarding the values of $d$ and the exponents $a$ and $c$, the following considerations apply:

(1) At La = La$_{Hinze}$, the probability distribution abruptly changes the power-law dependence to a new exponent theoretically equal to $-10/3$ \citep{Clarke2003}. Fitting the power law to the data, the local shape factor $d$ should be around 3, and the exponent $c$ approximately equal to $-4$ (close but different from $-10/3$ for $R_o/\langle R_o\rangle \lesssim 40$, see figure \ref{f11}). Interestingly, one has that La$_{Hinze}$ is very close to La$_{max}\simeq 2\times 10^5$, the maximum value of La for which jet droplets are ejected, and therefore the exponent $c$ plays a secondary role in the global aerosol distribution since film droplets are much smaller than jet droplets in this size range.

(2) The uncertainty of measurements at the Aitken mode range, or the sheer absence of data make a precise determination of the shape of the distribution function and the calculation of the exponent $a$ challenging for $R_o/\langle R_o\rangle < 0.1$, if not useless. This is marked as a dotted line in figure \ref{f11}. Since no droplets are ejected for La $<$ La$_{min}$, the choice of La$_{min}$ as the turning point where the bubble size distribution decays, at least from the droplet generation side, is consistent. The calculations for the global aerosol distribution are insensitive to the $a$ values as long as the decay exponent $a$ is steeper than 3 for $R_o/\langle R_o\rangle \rightarrow 0$.

An important question raised by \cite{Neel2021} is whether the actual probability distribution of bubbles popping at the surface of seawater is reproduced by the probability distribution measured in the bulk, given by (\ref{SMD}). While a one-to-one correspondence was assumed in \cite{Berny2021} and in \cite{Jiang2022}, given the very different raising velocities of the bubble size spectrum and the stages of development of the breaking wave, that assumption is called into question. However, it can be sustained as the most statistically consistent one for the purposes of this study since the global ensemble distribution of ejected droplet radii must consider the presence of bubbles, transient cavities, liquid ligaments, films and spumes necessarily making liquid-gas surfaces present at all scales in the turbulent motion, from about $R_o\sim l_\mu$ to about several centimeters (i.e. more than six orders of magnitude). Thus, bulk bubbles capable of generating jet droplets are actually exposed to liquid surfaces in a turbulent ocean much more frequently than the tranquil raising of bubbles in a laboratory tank.

\subsection{Ensemble droplet size statistics}\label{sec8}

Once the droplet size statistics, the number of ejected droplets per bursting for each droplet generation mechanism, and the statistics of bubbles are quantitatively described, one can easily calculate the theoretical ensemble probability $P(\chi_d)$ of a given ejected droplet size $\chi_d=r_d/l_\mu$.

The non-dimensional ensemble probability can be understood as the expectancy of the number of droplets $n_d$ as a function of La = $R_o/\mu$ under the combined probability of the variable La (the non-dimensional bubble radius) given by $\theta($La$)$ and the probability of $\chi_d$ for the average droplet radius $\langle \chi_d \rangle=\langle r_d \rangle/l_\mu$, which is a function of La as well \citep{Lhuissier2012,Berny2021}:
\begin{equation}
    P(\chi_d)= \int_{0}^{\infty} \theta(\text{La}) \frac{p\left(r_d/\langle r_d\rangle \right)}{\langle r_d \rangle/l_\mu} n_d\,  d \text{La}=
   \int_{0}^{\infty} \theta(\text{La}) \frac{p\left(\chi_d/\langle \chi_d\rangle \right)}{\langle \chi_d \rangle} n_d\,  d \text{La},
    \label{P}
\end{equation}
where $p(y=\chi_d/\langle \chi_d\rangle)$ is the Gamma distribution and $\theta($La$)$ is the p.d.f. of the bubble radius in the liquid bulk beneath the average position of the turbulent liquid free surface. The global average jet droplet radius $ \langle\langle \chi_d \rangle\rangle=\langle\langle r_d\rangle\rangle/l_\mu$ is simply:
\begin{equation}
    \langle\langle \chi_d \rangle\rangle = \int_0^\infty \chi_d P(\chi_d) d \chi_d.
\end{equation}
Note that the average $\langle \chi_d\rangle$ of the stochastic variable $\chi_d$ in a single bursting event is a function of La and Mo only, given by the same expression (\ref{RRo}) as that for the value of $\chi_d$ for the first ejected droplet (i.e. $R/l_\mu$), but with $\alpha_1$ as a {\sl free parameter} depending on the environment. This free parameter should be universal for seawater in air at average ocean atmospheric conditions (pressure and temperature).

Observe that the integration of (\ref{P}) is performed on the complete La domain and the kernel vanishes at both La $\rightarrow 0$ and $\infty$. In contrast, the kernel of the aggregated distribution in \cite{Berny2021} diverges for La $\rightarrow 0$: the shape of the aggregated distribution is therefore strongly determined by the limits of integration in that work.

\section{Results and Discussion}\label{sec9}

An inventory of well established data sets of the aerosol concentration spectra from the extensive literature reporting atmospheric aerosol measurements from the ocean is selected, including ultrafine particles in the Aitken and accumulation modes \citep{Pohlker2021}, CCN, and INP. Selection of measurements is made at or around the marine boundary layer (MBL), approximately at the average ocean temperature (15 $^o$C), or in laboratory measurements where the described conditions reasonably reproduce the open ocean ones \citep{ODowd1997,Hoppel2002,Clarke2003,Martenson2003,Sofiev2011,Gras2017,Wang2017,Erinin2019}. Also, for completeness, the numerical simulation data from \cite{Berny2021} for bubble swarms is included. Given the width of the aerosol size range (about five orders of magnitude), the ranges of validity of the measurement instruments, the physical effects influencing their performance or the aerosol concentrations measured, and the treatment of samples should be appropriately addressed to build a reasonable overall experimental p.d.f. Obviously, the number concentrations provided by published measurements should be scaled to obtain probability measures. In effect, the collected experimental data from the literature \citep{ODowd1997,Deane2002,Hoppel2002,Clarke2003,Clarke2006,Sofiev2011,Quinn2015,Wang2017,Erinin2019,Berny2021}) are scaled according to the procedures described in the Appendix \ref{A2} to build an experimental probability density function (pdf) $P(\chi_d=r_d/l_\mu)$. The matching of the experimental pdf shape at the overlapping ranges among the different data sets provides a good measure of reliability.

\begin{figure*}[!t]%
\centering
\includegraphics[width=0.85\textwidth]{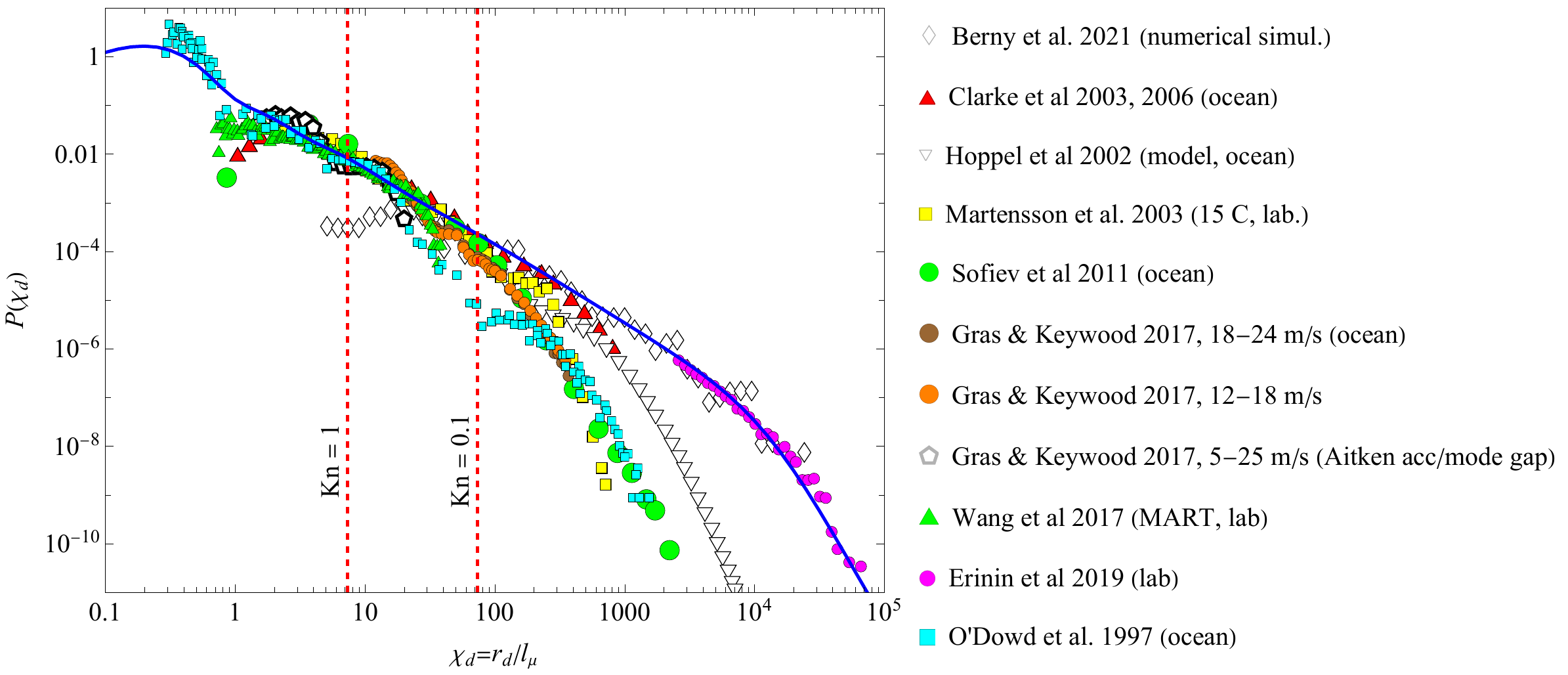}
\caption{The probability distribution function $P(\chi_d)$ for the radius $\chi_d=r_d/l_\mu$ of the total ensemble of ejected droplets from the sea at an average temperature of $T=15^o$C compared to measurements of SSA and OA from different authors. We also note where the measurements were performed for each data set. The pdf $P(x)$ and the resulting overall average droplet radius $\langle\langle r_d\rangle\rangle \simeq 0.4\, \mu$m is calculated with $\alpha_1=0.001$. The finest aerosol size range described by \cite{ODowd1997} could be due to the smallest relics of jet nano-droplets on which VOCs and other vapors condensate.}
\label{f12}
\end{figure*}

The calculations include an approximate reconstruction of the spray radii from which the given aerosol sizes originate. Given the focus of this work on CCN and INP, their size range is used as the main overlapping range among the different data sets to re-scale the number concentrations and to establish a reasonably continuous pdf. The upper envelope of the experimental data sets is the reference line reducible to a probability distribution. The result is plotted in figure \ref{f12}. Obviously, the requirement of $P(\chi_d)$ being a probability distribution determine the scaling of the different experimental data sets.

A remarkable overall fitting to the proposed model is achieved for $\alpha_1=10^{-3}$, with $\langle\langle r_d\rangle\rangle \simeq 0.4\, \mu$m, corresponding to a nascent SSA of avearge diameter $\langle\langle D_p\rangle\rangle =200$ nm. The sensitivity of the model to the main parameter $\alpha_1$ is given in the Appendix \ref{A3}. The result of the model fitting points to an extremely large droplet generation in the nanometric range. This implies the disintegration of extremely thin, nanometric-sized liquid ligaments ejected at enormous speeds (four to five times the average one of air molecules at standard atmospheric conditions) from bursting bubbles from about 5 to 30 micrometers into several thousands of droplets in the range from about 5 to 20 nm. This fits reasonably well the extreme ultrafine SSA size spectrum of \cite{ODowd1997} for typical maritime North East Atlantic measured number distribution using SMPS. This measured SSA range from about 4 to 15 nm would originate from seawater droplets with radii around 0.4 to 1.2 times the natural length $l_\mu$ (19.5 nm for seawater at 15$^o$C), the size range where the liquid jet acquires its maximum speeds of 4 to 15 times the natural velocity $v_\mu$ (see \cite{GC2021}, page A12-7, figure 2). To indicate the gas flow regimes, red dashed vertical lines in figure \ref{f12} indicate the ordinate $\chi_d$ values for Kn = 1 and 0.1; between these lines, one has rarefied flow of droplets in the air. At the left of Kn = 1, one has free molecular flow regime.

\conclusions  

The main conclusion from this work, on the basis of available evidences, are:

(1) When constructing an ensemble probability density function of the measured oceanic aerosol sizes, film droplet production alone, including the very recently proposed film flapping mechanism \citep{Jiang2022}, appears to fall orders of magnitude short of explaining the actual numerical concentration of measured aerosols in the submicron range. First, the energy densities entailed in the liquid film dynamics is insufficient to reach scales comparable or below the natural length scale of seawater, $l_\mu=\mu^2/(\rho \sigma) = 19.5$ nm. And secondly, the number of droplets produced by the film bursting mechanisms decrease drastically as the bubble radius $R_o$ decrease below 200 $\mu$m.

(2) In contrast, not only the bubble size distribution strongly favors bubble sizes between 5 and 200 $\mu$m, but also this bubble size range generates a large number of jet droplets in the submicron range per bursting event. In effect, the astonishing concentration of kinetic energy per unit volume at the point of collapse for bubble sizes around tens of microns is sufficient to foster jetting scales much smaller than $l_\mu$.

These conclusions would imply a complete reconsideration of the aerosol production physical paths from the ocean: the vast majority of these aerosols would have their elusive birth in the uterus-like nano-shape (figure \ref{f6}) of the bursting bubble at the very latest instants of collapse, with enormous implications on the understanding of oceanic aerosols and their origin. Naturally, the statistical agreement shown is a necessary but not sufficient condition for an irrefutable attribution to the described mechanisms. However, no other mechanisms have been described so far at this level of accuracy to explain the origin of submicrometer and nanometer scale nuclei (cloud condensation nuclei, ice nucleation particles, volatile organic compound nuclei, etc.) of primary and secondary oceanic aerosols, and to explain their high diffusivity from the ocean surface. More importantly, the accuracy of this model is such that it provides an optimal component for ocean aerosol fluxes in global climate models, in the absence of a better solution.


%
%
%
%
%
%
%
%

\appendix
\section{Experimental data reliability and spectral limits}
\label{A1}

The spray (primary marine aerosol) size spectra measured in the collection of selected data sets span five orders of magnitude, from ultrafine to coarse particles. This demands the use of measuring instruments based on different technologies: \citep{Clarke2003,Clarke2006,Martenson2003,Sofiev2011,Gras2017,Wang2017} have used ultrafine condensation particle counters (e.g. TSI3760, TSI3787, TSI3010, etc.), mobility analyzers (e.g. TSI3081, TSI3790), aerodynamic particle sizers (e.g. TSI3321), automated static thermal gradient cloud chamber, or active scattering aerosol spectrometer probe (ASASP-X), each of them with a relatively narrow measuring range capability compared to the whole marine spray size span. Consequently, open ocean measurements using these instruments tend to underestimate significantly the actual content of the spray spectra in the air layers in contact with the sea surface, since the size spectrum beyond the accumulation mode tends to settle. The only data correctly representing the actual spray size contents in that region are those from \citep{Berny2021}, with a direct account -drop by drop- from simulations, and from \citep{Erinin2019}.

\section{Data treatment}
\label{A2}

Reported SSA particle sizes are scaled with $l_\mu$ measured at the average ocean temperature of 15$^o$C. Most authors report the values of dry aerosols (obtained with a variety of drying means and temperatures from ambient ones to about 300$^o$C), except Eirin et al. \citep{Erinin2019} and Berny et al. \citep{Berny2021}). Besides, while \citep{ODowd1997}, \citep{Hoppel2002}, \citep{Erinin2019}, and \citep{Berny2021} report the droplet or aerosol radius, the rest give the diameter. The reconstruction of the corresponding droplet size from dry residues is made using a standard 3.5\% salt concentration in the ocean when the drying is considered complete. Naturally, the concentration $N$ per unit volume reported by most authors, which is strongly dependent on the actual local conditions of measurements (fundamentally wind and wave amplitude \citep{Deane2002,Clarke2003,Clarke2006,Sofiev2011,Quinn2015,Wang2017}) is scaled to represent a true p.d.f. under the hypothesis that each reported concentration (their ordinates $N$) correctly describes at least a fraction of the complete number density spectrum.

\section{Model sensitivity to the free parameter $\alpha_1$}
\label{A3}

\begin{figure}[!t]%
\centering
\includegraphics[width=0.50\textwidth]{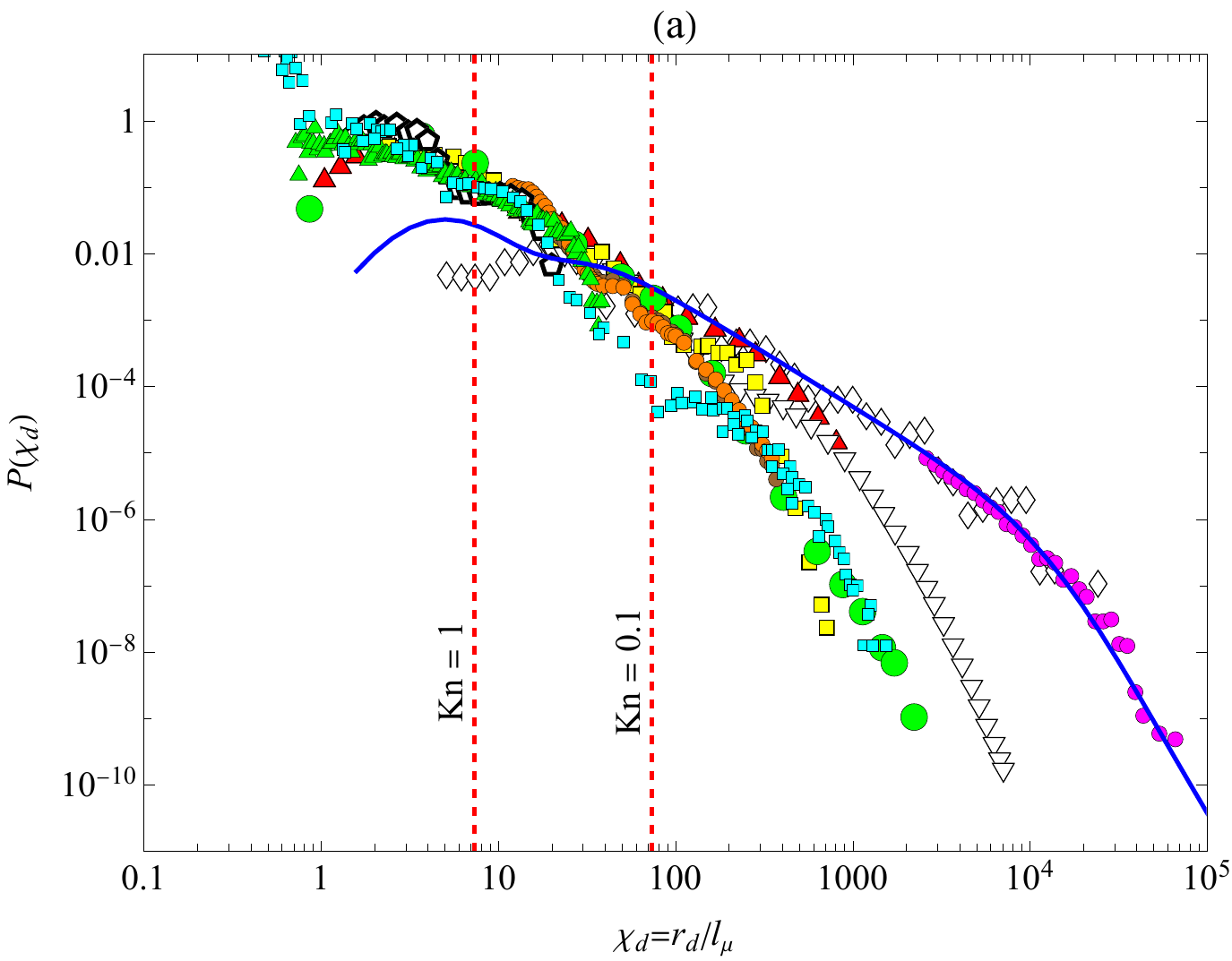}
\includegraphics[width=0.50\textwidth]{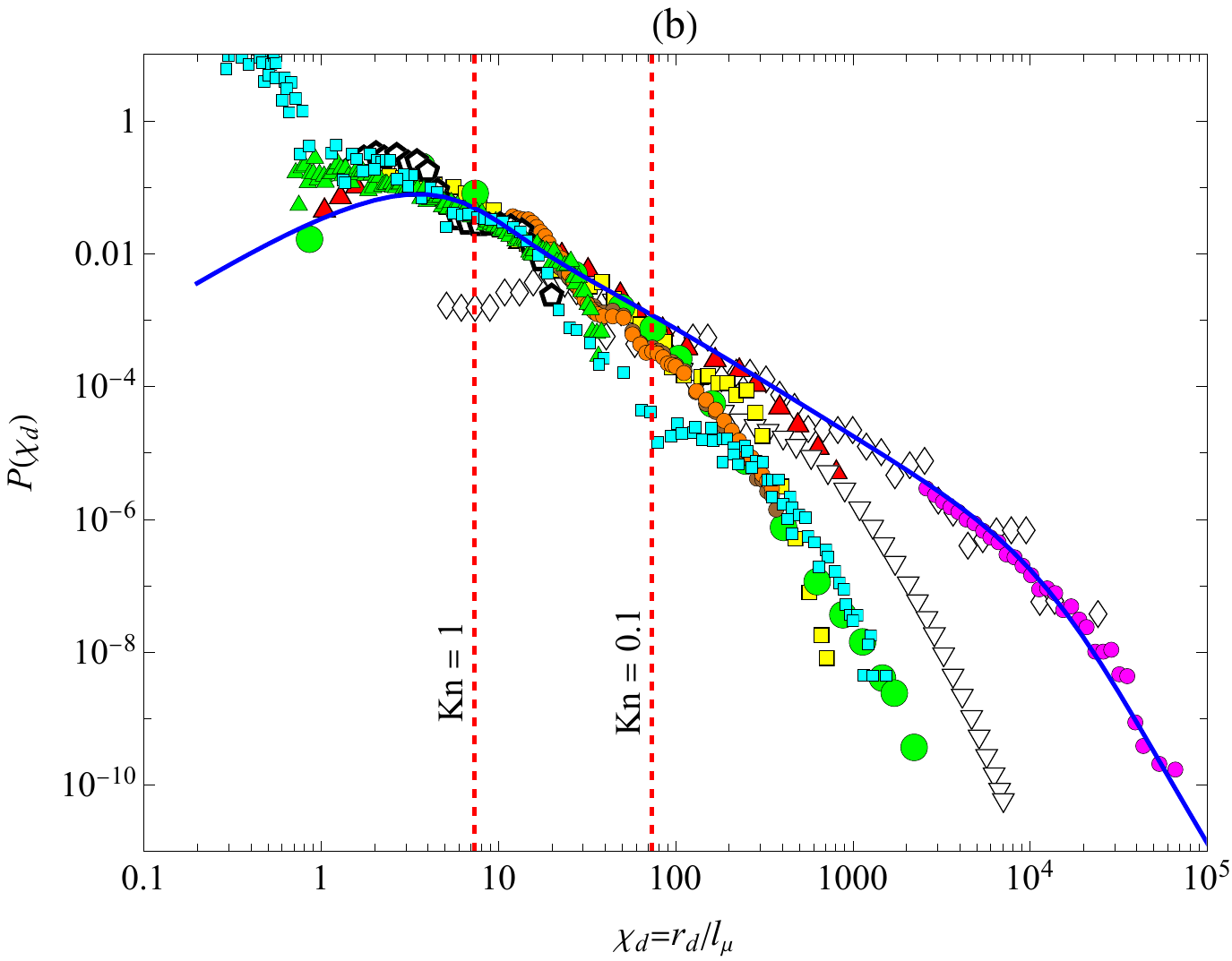}
\caption{The probability distribution function $P(\chi_d)$ for the radius $\chi_d=r_d/l_\mu$ of the total ensemble of ejected droplets from the sea at an average temperature of $T=15^o$C compared to measurements of SSA and OA from different authors. We also note where the measurements were performed for each data set. The p.d.f. $P(x)$ and the overall average droplet radius $\langle\langle r_d\rangle\rangle$ is calculated with $\alpha_1=$ (a) 0.13 \citep{GC2021}, and (b) 0.01. The corresponding averages $\langle\langle r_d\rangle\rangle$ are 5.3 and 1.3 $\mu$m. Note that the ordinate values are automatically adjusted in each plot. Plot markers correspond to those of figure \ref{f12}, main text.}
\label{f13}
\end{figure}

Figure \ref{f13} shows the fitting of the model given by equation (11), main text, to the data sets, under different hypotheses:

(1) Figure \ref{f13}(a): The model uses the fittings constants published in \citep{GC2021} ($\alpha_1=0.133$ and Oh$_c=0.03$, or La$_c=1111$), obtained from available data on individual BB using several different liquids which in the range of Laplace numbers yield measurable ejections. Naturally, these ejections happen in air at atmospheric conditions and their characteristic speeds are sufficiently smaller than the speed of sound to assume incompressibility throughout the whole bursting and ejection events. Note that the model perfectly fits the numerical simulation data from \citep{Berny2021}, made under these hypotheses (see figure 5, main text). The contribution of the flapping droplets \citep{Jiang2022} would be maximum in the range of CCN marked for this value of the free parameter ($\alpha_1=0.133$).

(2) Figure \ref{f13}(b): This intermediate model prediction uses an intermediate fitting ($\alpha_1=0.01$) between those which fits the numerical simulations of \citep{Berny2020} ($\alpha_1=0.032$) and from \citep{GC2021} ($\alpha_1=0.003$), see figure 5, main text. With some caveats, $\alpha_1=0.01$ would fit nearly all data sets, except the extreme ultrafine spectrum measured by O'Dowd at al. \citep{ODowd1997}. The contribution of flapping droplets would completely vanish compared to that of jet droplets for this $\alpha_1$ value.

The bubbles producing the size range predicted in figure 10, main text, should be present in surface seawater fully saturated (or supersaturated) with air. These conditions are indeed met under continuous wave breaking \citep{Wang2017,Deike2022}: due to their large internal air pressure, they should diffuse air into the surrounding water very effectively. In addition, each bursting bubble in the range from 10 to 500 micrometers can generate tiny bubbles that get trapped at the bottom of the cavity in the liquid \citep{Duchemin2002,Krishnan2017,GC2021}, which increases the air supersaturation at the surface microlayer.

\noappendix       




%



\competinginterests{The author report no conflict of interest.} 


\begin{acknowledgements}
This research has been supported by the Spanish Ministry of Economy, Industry and Competitiveness (Grants numbers DPI2016-78887 and PID2019-108278RB), and by Junta de Andaluc\'{\i}a (Grant number P18-FR-3623. Data from numerical simulations made by Jos\'e M. L\'opez-Herrera, already reported in \citep{GC2021}, are used in this work (figure \ref{f7}), which is especially acknowledged. Pascual Riesco-Chueca made very useful suggestions. Cristina de Lorenzo read the paper carefully and provided insightful comments.
\end{acknowledgements}

\end{document}